\documentclass[twocolumn]{aastex63}

\usepackage{apjfonts}
\usepackage{graphicx}
\usepackage{mathpazo}
\usepackage{amsmath}

\newcommand{\beq}{\begin{equation}}
\newcommand{\eeq}{\end{equation}}
\newcommand{\beqn}{\begin{eqnarray}}
\newcommand{\eeqn}{\end{eqnarray}}
\newcommand{\eq}[1]{Eq.~(\ref{#1})}
\newcommand{\eqs}[2]{Eqs.~(\ref{#1})~and~(\ref{#2})}
\newcommand{\kderiv}[1]{\frac{{\mathcal D} #1}{{\mathcal D}t}}
\newcommand{\advec}{\v{u}\cdot\del}
\newcommand{\St}{\mathrm{St}}
\newcommand{\xtimes}[2]{#1\times{10^{#2}}}
\renewcommand{\v}[1]{{\boldsymbol{#1}}}
\newcommand{\del}{\v{\nabla}}
\newcommand{\grad}{\del}
\newcommand{\Div}{\del\cdot}
\newcommand{\Laplace}{\nabla^2}
\newcommand{\hatz}{\hat{\v{z}}}
\newcommand{\haty}{\hat{\v{y}}}
\newcommand{\hatx}{\hat{\v{x}}}
\newcommand{\app}[1]{Appendix~\ref{#1}}
\renewcommand{\fig}[1]{Fig.~\ref{#1}}
\renewcommand{\table}[1]{Table~\ref{#1}}
\newcommand{\sect}[1]{Sect.~\ref{#1}}

\usepackage[encapsulated]{CJK}
\newcommand{\cntext}[1]{\begin{CJK}{UTF8}{bkai}#1\ignorespacesafterend\end{CJK}}

\newcommand{\rhoi}{\rho_i}
\newcommand{\rhos}{\rho_s}

\newcommand{\mpluto}{$M_{\rm Pluto} \ $}
\newcommand{\mplutoc}{$M_{\rm Pluto}$}

\def\apj{\rm ApJ}
\def\apjl{\rm ApJL}
\def\aj{\rm AJ}
\def\mnras{\rm MNRAS}
\def\nat{\rm Nature}
\def\aap{\rm A\&A}
\def\araa{\rm ARA\&A}
\def\icarus{\rm Icarus}
\def\planss{\rm Planet. Space Sci.}

\submitjournal{PSJ}

\shorttitle{Densities of Kuiper Belt objects}
\shortauthors{Ca\~nas et al.}

\begin{document}

\title{A solution for the density dichotomy problem of
  Kuiper Belt objects with multi-species streaming instability and
  pebble accretion}

\correspondingauthor{Wladimir Lyra}
\email{wlyra@nmsu.edu}

\author[0009-0000-4011-8677]{Manuel H. Ca\~nas}
\affiliation{New Mexico State University, Department of Astronomy, PO Box 30001 MSC 4500, Las Cruces, NM 88001, USA}

\author[0000-0002-3768-7542]{Wladimir Lyra}
\affiliation{New Mexico State University, Department of Astronomy, PO Box 30001 MSC 4500, Las Cruces, NM 88001, USA}

\author[0000-0001-6259-3575]{Daniel Carrera}
\affiliation{Department of Physics and Astronomy, Iowa State University, Ames, IA, 50010, USA} 

\author[0000-0001-7671-9992]{Leonardo Krapp}
\altaffiliation{51 Pegasus Fellow}
\affiliation{Department of Astronomy and Steward Observatory, University of Arizona, Tucson, Arizona 85721, USA}

\author[0000-0003-0801-3159]{Debanjan Sengupta}
\affiliation{New Mexico State University, Department of Astronomy, PO Box 30001 MSC 4500, Las Cruces, NM 88001, USA}

\author[0000-0002-3771-8054]{Jacob B. Simon}
\affiliation{Department of Physics and Astronomy, Iowa State University, Ames, IA, 50010, USA}

\author[0000-0001-5372-4254]{Orkan M. Umurhan}
\affiliation{NASA Ames Research Center, Space Sciences Division, Planetary Sciences Branch, Moffatt Field, CA 94035, USA} 

\author[0000-0003-2589-5034]{Chao-Chin Yang (\cntext{楊朝欽})}
\affiliation{Department of Physics and Astronomy, University of Alabama, Box 870324, Tuscaloosa, AL 35487-0324} 

\author[0000-0002-3644-8726]{Andrew N. Youdin}
\affiliation{Department of Astronomy and Steward Observatory, University of Arizona, Tucson, Arizona 85721, USA} 
\affiliation{The Lunar and Planetary Laboratory, University of Arizona}

\begin{abstract}
Kuiper belt objects show an unexpected trend, whereby large bodies have increasingly higher densities, up to five times greater than their smaller counterparts. Current explanations for this trend assume formation at constant composition, with the increasing density resulting from gravitational compaction. However, this scenario poses a timing problem to avoid early melting by decay of $^{26}$Al. We aim to explain the density trend in the context of streaming instability and pebble accretion. Small pebbles experience lofting into the atmosphere of the disk, being exposed to UV and partially losing their ice via desorption. Conversely, larger pebbles are shielded and remain more icy. We use a shearing box model including gas and solids, the latter split into ices and silicate pebbles. Self-gravity is included, allowing dense clumps to collapse into planetesimals. We find that the streaming instability leads to the formation of mostly icy planetesimals, albeit with an unexpected trend that the lighter ones are more silicate-rich than the heavier ones. We feed the resulting planetesimals into a pebble accretion integrator with a continuous size distribution, finding that they undergo drastic changes in composition as they preferentially accrete silicate pebbles. The density and masses of large KBOs are best reproduced if they form between 15 and 22\,AU. Our solution avoids the timing problem because the first planetesimals are primarily icy, and $^{26}$Al is mostly incorporated in the slow phase of silicate pebble accretion. Our results lend further credibility to the streaming instability and pebble accretion as formation and growth mechanisms. 
\end{abstract}


\section{Introduction}

The formation of kilometer-sized planetesimals from $\mu$m-sized dust particles is a process still relatively poorly understood, despite significant advances \citep{ChiangYoudin10, Johansen+14, Lesur+22}.
Sub-$\mu$m interstellar grains grow by collisional coagulation \citep{Safronov72,Nakagawa+81,Tominaga+21}, but accumulated evidence from laboratory experiments \citep{BlumWurm08,Guttler+10}, numerical simulations \citep{Guttler+09,Geretshauser+10,Zsom+10} and observations \citep{Perez+15,Tazzari+16,Carrasco-Gonzalez+19} suggest that growth is inefficient beyond millimeter (mm) and centimeter (cm) size (hereafter called ``pebbles'') due to bouncing, fragmentation, and drift \citep{DullemondDominik05,Brauer+08,Krijt+15}.  Therefore, an alternative route to planetesimal formation is needed, perhaps via direct gravitational collapse of pebble clouds \cite{YoudinShu02}.

A promising mechanism to overcome collisional barriers, and trigger gravitational collapse, is the streaming instability \citep{YoudinGoodman05,YoudinJohansen07,JohansenYoudin07}. In this feedback mechanism, particle overdensities trigger a gas flow that enhances the particle overdensities \citep{YoudinGoodman05,SquireHopkins20}. In the typical conditions of gas-rich protoplanetary disks, the overdensities caused by streaming instability have been shown to be massive enough to result in the formation of gravitationally bound objects \citep{Johansen+07, YangJohansen14, Carrera+15, Simon+16, Simon+17, Schafer+17, Yang+17, Li+19, Abod+19, Nesvorny+19, LiYoudin21}. Growth to planetary masses then continues via accretion of pebbles, as they lose energy from aerodynamical drag while passing a protoplanet, significantly increasing its effective accretional cross-section  \citep{Lyra+08b, JohansenLacerda10, OrmelKlahr10, LambrechtsJohansen12,Lyra+23}.

\begin{figure}
    \begin{center}
    \resizebox{\columnwidth}{!}{\includegraphics{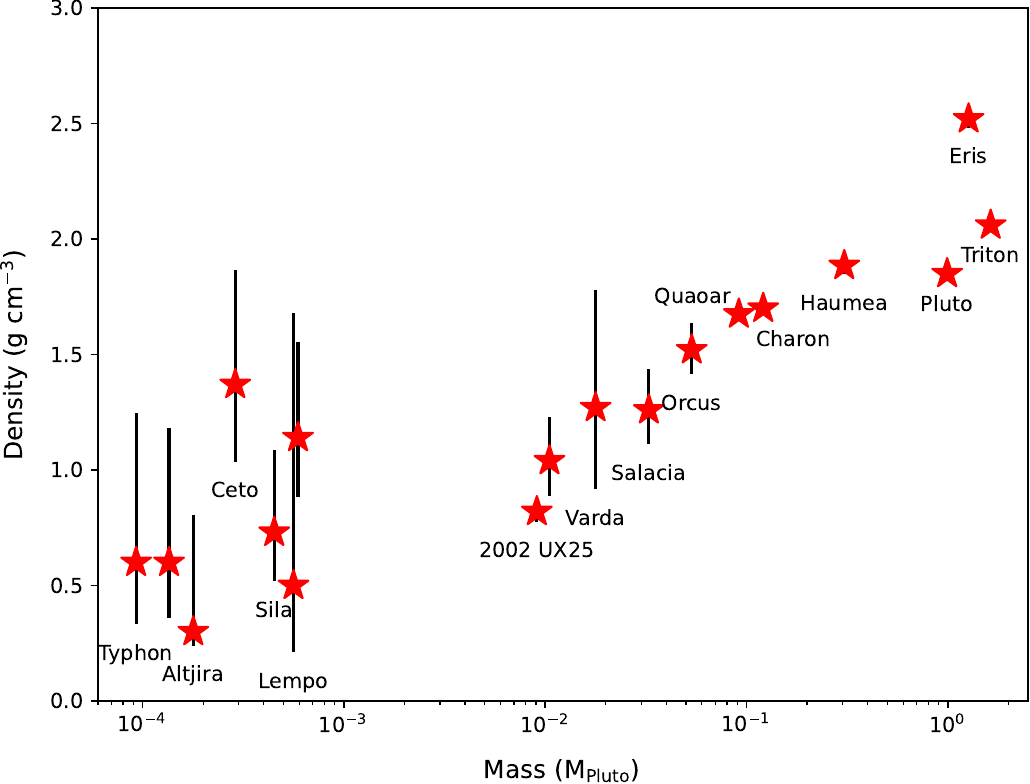}}
    \end{center}
    \caption{Density distribution of Kuiper belt objects. Small KBOs have density around 0.5\,g\,cm$^{-3}$ (with some dispersion), implying porosity, whereas larger bodies have increasingly higher density. Values from \citet[][and references therein]{BiersonNimmo19}, \citet{Morgado+23}, and \citet{Pereira+23}.} 
\label{fig:kbos_w_label}
\end{figure}

Given that the current models of planet formation involve the concentration and accumulation of neighboring pebbles, it is surprising that Kuiper belt objects (KBOs) demonstrate a large range of densities (from 0.5 g\,cm$^{-3}$ to 2.6 g\,cm$^{-3}$, \citealt{Brown13}), with a trend that the small KBOs ($\lesssim 2\times10^{-2}$ Pluto masses) have a near-constant density of $0.5$ g\,cm$^{-3}$ and larger KBOs have an increasingly larger density that reaches $\approx 2.5 \text{ g cm}^{-3}$ \citep{Stansberry+06, BrownButler07, Grundy+07, Brown12,Brown13,Stansberry+12, BarrSchwamb16,Ortiz+17, Grundy+19}. We show in \fig{fig:kbos_w_label} the mass vs density relation; the data for mass and density are from \citet[][and references therein]{BiersonNimmo19}, except for Quaoar, for which we use the more recently evaluated mass of $\xtimes{\left(1.20\pm 0.05\right)}{24}$\,g from \citet{Morgado+23} and \citet{Pereira+23}. 

One plausible explanation is that smaller KBOs are more porous, and as a KBO grows, porosity is removed through gravitational compaction. Assuming water ice (density $\approx$1\,g\,cm$^{-3}$) is the lowest density material of substantial abundance in the composition of the small objects, the bulk density of 0.5 g\,cm$^{-3}$ implies a porosity of at least 50\% for them. The larger objects (e.g. Eris, Pluto, and Triton) should be less porous even in the unlikely case of 100\% rocky composition. Interestingly, \citet{Baer+11} find a similar trend in the asteroid belt, with the largest asteroids showing porosities less than 20\%, with an abrupt change for diameters under 300\,km where a range of porosities from 0-70\% is seen.  

Gravitational compaction at constant composition was explored by \citet{BiersonNimmo19}. Assuming a rock mass fraction of 70\%, these authors successfully match 15 of 18 KBOs within two standard deviations allowed by their model (or 11 within one standard deviation). While promising, this scenario poses a timing problem requiring KBOs to be formed at least four million years after the formation of the Calcium-Aluminum-rich Inclusions (CAIs). The reason for this time constraint is that heat from the decay of the short-lived radioactive isotope $^{26}$Al, with a half-life of 0.7\,Myr \citep{Norris+83}, would act to melt the objects and remove porosity. Thus, if KBOs formed early in the solar nebula with a high rock fraction, we would expect to see high-density, low-mass KBOs, which are not seen. 

The required time delay, however, is unlikely because four million years is potentially longer than the disk lifetime \citep[e-folding time 2.5 Myr, e.g.][]{Ribas+14}, which would contradict evidence for KBO formation by gravitational collapse in a gas disk \citep{Nesvorny+10,Nesvorny+19,Lisse+21}, and Arrokoth needing nebular drag for the two lobes to come into contact \citep{Mckinnon+20,Lyra+21}. The special timing also contradicts indication that planets might form early in protoplanetary disks \citep[e.g.][]{ALMA+15,Manara+18,Tobin+20,ediskIV,ediskV}.  Thus we search for another explanation for the density trend, where the large change in density stems from compositional differences between large and small KBOs, with large KBOs containing a higher rock fraction than their smaller counterparts. 

We consider formation beyond the water ice line, the region of the disk where it is cold enough for water to condense into solid ice grains. While coagulation of ices and silicates beyond the snowline should produce pebbles of homogeneous composition, a compositional difference in pebbles should be expected from preferential depletion of ices in small grains. Due to turbulence and bulk vertical motions in the disk -- caused by, e.g., the vertical shear instability \citep[VSI,][]{Nelson+13,LinYoudin15,Lesur+22} or the magnetorotational instability \citep[MRI,][]{BalbusHawley98,Lyra+08a,Johansen+09, BaiStone11,Simon+18} if the disk is sufficiently ionized --, smaller grains are lofted up in the atmosphere of the disk. There, they are exposed to stellar UV irradiation, which causes removal of ice by photodesorption \citep[photosputtering, ][]{Westley+95a,Westley+95b,Bergin+95,Andersson+06,AnderssonDishoeck08,Oberg+09, Krijt+16,Krijt+18,Potapov+19,Powell+22}. In an optically thin environment, the removal rate produced by solar UV irradiation is estimated at 4\,mm per Myr \citep{HarrisonSchoen67} at 10\,AU (similar to the coagulation rate at steady state), obliterating small icy grains and perhaps explaining their apparent absence in debris disks \citep{Grigorieva+07,StuberWolf22}. Evidence for this process in a gas-rich disk is seen in the presence of water vapor at very low temperatures in circumstellar disks \citep{Hogerheijde+11}, which is interpreted as water molecules released by non-thermal desorption into the gas phase, before quickly recondensing as amorphous ice \citep{Ciesla14}.

The height at which the disk becomes optically thick to UV is estimated at 3 scale heights for a primordial disk where all the dust is in sub-$\mu$m size \citep{Krijt+18}. Yet, the optical depth at a given height decreases with time as the sub-$\mu$m grains responsible for the opacity coagulate into larger grains that settle toward the midplane. As most of the solid masses is converted into larger grains, the mass remaining in sub-$\mu$m grain species decreases significantly. When the dust-to-gas ratio of these sub-$\mu$m grain species decreases to $10^{-4}$, typical of Class I-II disks, the gas becomes optically thin to UV as close to the midplane as 1.5 scale heights, a layer where even low levels of turbulence would suffice to loft sub-mm sized grains into. Bulk motions due to the VSI would make it even easier to get these particles to the UV-irradiated layer, and cosmic ray desorption \citep{Silsbee+21,Sipila+21,Rawlings22} would further enhance the loss of ice coating from grains. As a result of these processes, smaller grains should have less ice than larger grains that reside near the disk midplane, shielded from UV and cosmic rays. A bimodal distribution of grains is then expected, with smaller ice-poor pebbles (hereafter ``silicate'' pebbles), and larger pebbles composed of dirty ice (hereafter ``icy'' pebbles).

Considering the results of \citet{DrazkowskaDullemond14,Carrera+15,Yang+17}, and \citet{LiYoudin21}, where the effectiveness of the streaming instability is tested amongst various grain sizes, we expect the (larger) icy pebbles to be more conducive to the streaming instability, while the (smaller) silicate pebbles are more tightly coupled to the gas and hence participate less \citep[see also ][]{YangZhu21}. The outcome is that planetesimals formed beyond the ice line should be mostly icy. If this hypothesis is correct, there remains the question of how objects in the Kuiper belt achieve densities beyond $2.0 
\text{ g cm}^{-3}$. The answer might be through the subsequent processes of pebble accretion and compaction. Pebble accretion has a strong dependency on both grain size and planetesimal mass \citep{Lyra+23} such that, as a planetesimal grows, the favorable grain size for pebble accretion also changes. This allows protoplanets to accrete a wealth of silicates through pebble accretion. The accretion of silicates, in combination with limited removal of porosity from gravitational compaction, would result in a natural trend where low-mass objects (i.e., those that failed to accrete pebbles) have low densities, and the bodies increase in density as a higher fraction of their mass comes from pebble accretion. This is the scenario we explore in this work.

This paper is organized as follows. In \sect{sect:model} we discuss our model, including the hydrodynamical
simulation of the streaming instability and the numerical integration
of polydisperse pebble accretion. In \sect{sect:results} we describe our results of both the
streaming instability and pebble accretion. In \sect{sect:discussion} we discuss
the interpretation of our results and compare it with other works that
sought to explain the density trend of KBOs. In \sect{sect:limitations} we discuss the limitations of the model and future works. We conclude the work in \sect{sect:conclusions}.

\begin{table*}
\caption{Symbols used in this work.}
\label{table:symbols}
\begin{center}
\begin{tabular}{lllclll}\hline
Symbol             & Definition                 & Description                           && Symbol            & Definition                & Description                              \\
\hline
  $t$              &                            & time                                  && $r_0$             &                           & arbitrary origin of shearing box         \\
  $x,y,z$          &                            & Cartesian coordinates                 && $p$               & \eq{eq:pressure}          & gas pressure                             \\
  $\varepsilon$    & $\rho_d/\rho_g$            & dust-to-gas density ratio             && $c_s$             & \eq{eq:cs}                & the sound speed                          \\
  $Z$              & $\frac{\int\rho_d dV}{\int \rho_g dV}$&dust-to-gas mass ratio      && $dV$              & $dx\,dy\,dz$              & volume infinitesimal                     \\
  $v_K$            & $\varOmega_0 r_0$          & Keplerian velocity                    && $\Delta v$        & $\eta v_K$                & gas velocity reduction                   \\
  $\gamma$         &                            & adiabatic index                       && $\tau$            & \eq{eq:tau}               & particle friction time                   \\
  $v_{\rm th}$     & $\sqrt{\frac{8}{\pi}} c_s$ & thermal velocity                      && $\rho_{\bullet}$  &                           & pebble material density                  \\
  $k_B$            &                            & Boltzmann constant                    && $\v{v}$           & $(v_x, v_y, v_z)$         & pebble velocity                          \\     
  $T$              &                            & temperature                           && $\lambda_{\rm SI}$& $\eta r$                  & streaming instability length scale       \\
  $m_H$            &                            & atomic mass unit                      && $H_d$             & \eq{eq:dust_scale_height} & pebble scale height                      \\
  $\varOmega_K$    & \eq{eq:kep_freq}           & Keplerian frequency                   && $N$               & $N_x, N_y, N_z$           & box resolution                           \\  
  $M$              &                            & stellar mass                          && $\widetilde{G}$   & \eq{eq:g_tilde}           & dimensionless gravitational constant     \\
  $\delta$         &                            & dimensionless pebble diffusion        && $\rho_d$          &                           & volume density of pebbles                \\
  $\Delta$         &                            & mesh spacing                          && $a$               &                           & pebble radius                            \\      
  $\phi$           &                            & porosity                              && $\rm{St}$         & $\varOmega_0\tau$         & Stokes number of pebbles                 \\
  $R$              &                            & protoplanet radius                    && $\varPhi$         &                           & self-gravitational potential             \\
  $\rhos$          &                            & silicate density                      && $\mu$             &                           & mean molecular weight                    \\
  $\rhoi$          &                            & ice density                           && $L$               & $L_x$=$L_y$=$L_z$         & length of shearing box sides             \\
  $H$              & $c_s/\varOmega_0$          & gas scale height                      && $\alpha$          & \eq{eq:alpha}             & dimensionless gas viscosity              \\
  $\v{u}$          & $(u_x, u_y, u_z)$          & gas velocity in Cartesian coordinates && $N_p$             &                           & number of numerical particles            \\
  $\varOmega_0$    &                            & Keplerian frequency at $r_0$          && $Q$               & \eq{eq:toomre_q}          & Toomre $Q$ parameter                     \\
  $\rho_g$         &                            & gas density                           && $\rho_R$          & \eq{eq:roche_density}     & Roche density                            \\
  $h$              & $H/r_0$                    & disk aspect ratio                     && $\dot{M}$         &                           & mass accretion rate                      \\
  $\eta$           & \eq{eq:eta}                & dimensionless pressure support        && $r$               &                           & astrocentric distance                    \\
  $u_0$            & $-\frac{3}{2}\varOmega_0 x$& Keplerian shear flow                  && $M_p$             &                           & protoplanet mass                         \\
  $\theta$         & \eq{eq:heaviside}          & Heaviside step function               && $\xi$             & \eq{eq:boxcar}            & boxcar function                          \\
  $G$              &                            & gravitational constant                && $\Pi$             & $\Delta v/c_s$            & dimensionless gas velocity reduction     \\
  $V$              &                            & protoplanet volume                    && $m_j$             &                           & mass of protoplanet in constituent $j$   \\
  $\rho$           &                            & protoplanet density                   && $V_j$             & $m_j/\rho_j$              & volume occupied by constituent $j$       \\  
  $F_j$            & $m_j/M_p$                  & mass fraction of constituent $j$      && $f_j$             & $V_j/V$                   & volume fraction of constituent $j$       \\ 
  $t_{\rm settle}$ & \eq{eq:settling}           & pebble settling time                  && [$^{\rm 26}$Al]   &                           & isotopic abundance of $^{\rm 26}$Al      \\
  $\mathcal{H}$    &                            & heat production rate per mass         && $c_p$             &                           & heat capacity of water ice               \\
  $t_{1/2}$        &                            & half-life of $^{\rm 26}$Al            && $\lambda$         & $\ln(2)/t_{1/2}$          & decay constant of $^{\rm 26}$Al          \\
\hline
\end{tabular}
\end{center}
\end{table*}

\section{The Model}
\label{sect:model}

We use the {\sc Pencil Code}\footnote{http://pencil-code.nordita.org} \citep[see][and references therein for details]{PencilCode} to model the formation of the first planetesimals by streaming instability of ices and silicates. We employ the shearing-box approximation \citep{Brandenburg+95,Hawley+95}, centered at an arbitrary position $r_0$ and orbiting at the corresponding Keplerian frequency $\varOmega_0$. Our model includes gas, solids, vertical stellar gravity, and dust self-gravity; the latter allows a concentration of solids to collapse into planetesimals once the Roche density is exceeded.

\subsection{Equations of motion}

We consider an isothermal gas disk such that any gain in internal energy is considered to be radiated away effectively. While the gas is solved on a uniform mesh, the pebbles are represented by Lagrangian particles. We ignore the selfgravity of the gas (explained a posteriori in \sect{sect:parameters}), but consider the gravity of the dust grains. Under this assumption, the equations of motion for the system are

\begin{eqnarray}
  \kderiv{\rho_g} &=& -\rho_g \Div \v{u}, \label{eq:gcont}\\
  \kderiv{\v{u}} &=& -\frac{1}{\rho_g}\grad{p} - 2\varOmega_0\hatz\times\v{u}+ \frac{3}{2} \varOmega_0 u_x\haty - \varOmega_0^2 \v{z} +\v{f}_g \nonumber\\
                     &&+ 2\varOmega_0 \Delta v\hatx, \label{eq:eqgas}\\
\frac{d\v{x}}{dt} &=& \v{v}  - \frac{3}{2}\varOmega x\haty,\label{eq:position-update}\\
\frac{d\v{v}}{dt} &=& -2\varOmega_0 \hatz\times \v{v} + \frac{3}{2}\varOmega_0 v_{x} \haty - \varOmega_0^2 \v{z} + \v{f}_{d} -\grad{\varPhi},\label{eq:eqdust}\\
    \Laplace{\varPhi} &=& 4\pi G \rho_d.\label{eq:poisson}
\end{eqnarray}

\noindent Here, $\rho_g$ is the gas density, $t$ is time, $\v{u}$ is the gas velocity, $p$ is the gas pressure, $\v{x}$ and $\v{v}$ are the position and velocity vectors of the pebbles, respectively, $\rho_d$ is the volume density of pebbles, $\varPhi$ is the selfgravity potential, $G$ is the gravitational constant, and ($x$,$y$,$z$) are the local Cartesian coordinates of the shearing box. The terms $f_d$ and $f_g$ are the drag force and its backreaction, respectively, explained in \app{app:dragforce}.

The quantity $\Delta v = \eta v_K$ is the orbital velocity reduction of the gas due to the pressure support, and $\eta$ is related to the global-scale radial pressure gradient \citep{Nakagawa+86},

\begin{equation}\label{eq:eta}
    \eta \equiv - \frac{1}{2\rho_g\varOmega_K^2r}\frac{\partial p}{\partial r}. 
\end{equation}

\noindent A table of symbols is provided in \table{table:symbols}. We have defined the operator

\begin{equation}
    \kderiv \ = \frac{\partial}{\partial t} + u_0 \frac{\partial}{\partial y} + \advec , 
\end{equation}

\noindent where $u_0 = -\left(3/2\right)\varOmega_{0}x$ is the linearized Keplerian shear flow. The pressure is related to the sound speed $c_s$ by the equation of state 

\begin{equation}\label{eq:pressure}
    p = \rho_g c_s^2/\gamma,
\end{equation}

\noindent which closes the system. Here $\gamma=1$ is the adiabatic index for an isothermal gas.  

The gas is initially in hydrostatic equilibrium between its own pressure and the vertical gravity from the central star. This results in the gas density having a vertical profile

\begin{equation}
\label{eq:gauss}
    \rho_g\left(z\right) = \rho_0 \exp\left(-\frac{z^2}{2H^2}\right),
\end{equation}

\noindent where $\rho_0$ is the midplane gas density and

\beq
H\equiv \frac{c_s}{\varOmega_0}
\label{eq:scaleheight}
\eeq

\noindent is the gas scale height.

\subsection{Box Domain and Resolution}\label{subsec:model_param}

We choose the length of our box such that the scales of the streaming instability, $\lambda_{\rm SI}\approx r\eta$, are well contained within the box. Given \eq{eq:scaleheight}, we can rewrite this length in terms of the scale height as $\lambda_{\rm SI} \approx hH$, where the disk aspect ratio $h \equiv H/r_0$. The sound speed is, for constant adiabatic index $\gamma$ and mean molecular weight $\mu$, solely dependent on temperature 

\begin{equation}\label{eq:cs}
    c_s^2 = \frac{\gamma k_B T}{\mu m_H}.
\end{equation}

\noindent We substitute the values $\gamma = 1$ for the adiabatic index of an isothermal flow, $\mu=2.3$ for the mean molecular weight (i.e., 5 $\rm{H}_{2}$ for every 2 He), $k_B$ and $m_H$ are the Boltzmann constant and atomic mass unit, respectively. The disk temperature $T$ is found using the temperature relation of \citet{ChiangGoldreich97}. In this model at $r=45\text{\,AU}$, the disk temperature is $T \approx 30\text{ K}$, which results in a sound speed of $c_s \approx 0.3\text{ km s}^{-1}$. The Keplerian frequency is

\begin{equation}\label{eq:kep_freq}
    \varOmega^2_K \equiv \frac{G M}{r^3},
\end{equation}

\noindent where $M$ is the mass of the host star, and $r$ is the astrocentric distance. At 45\,AU for a solar mass star, the Keplerian frequency is $\varOmega_K = 6.596 \times 10^{-10}\text{ s}^{-1}$ resulting in a scale height of $H \approx 3.3\text{\,AU}$. Thus, we have the characteristic length of the streaming instability being $\eta r \approx 0.07H$, placing an upper limit for the distance between mesh cells, $\Delta$. Even for low resolutions of 16 mesh cells along each axis, a cubic box of sides $L = 0.2H$ ($0.66\text{\,AU}$ for $H=3.3\text{\,AU}$) will ensure that the streaming instability is resolved. Consequently, this is the chosen size of our box.

While the size of our box is largely determined by $\lambda_{\rm SI}$, the required resolution of our mesh is primarily constrained by the dust scale height, $H_d$, which is determined as a balance between vertical stellar gravity and turbulent diffusion

\begin{equation}\label{eq:dust_scale_height}
    H_d = H\left(1+\frac{\St}{\delta}\right)^{-1 / 2},
\end{equation}

\noindent where $\delta$ is a dimensionless diffusion parameter \citep{Dubrulle+95, JohansenKlahr05, YoudinLithwick07,LyraLin13}  related to (in the case of isotropic turbulence, \citealt{Yang+18}) the dimensionless Shakura-Sunyaev $\alpha$ viscosity \citep{ShakuraSunyaev73}

\begin{equation}\label{eq:alpha}
    \alpha \equiv \frac{\langle \rho_g\, \delta v_x \, \delta v_y\rangle}{p}. 
\end{equation}

\noindent Here $\delta v_i$ is the turbulent deviation from the mean flow in the direction $i$. The $\alpha$-viscosity parameter 
is related to $\delta$ by \citep{YoudinLithwick07}

\begin{equation}
    \delta = \alpha\left(1 + \St^{2}\right)^{-1}.
\end{equation}

\eq{eq:dust_scale_height} is strictly valid only when particles are completely passive. While this is not the case for simulations of the streaming instability (as feedback is essential to this instability), we adopt this formula for simplicity. Also, \eq{eq:alpha} ignores magnetic stresses.

We do not consider external turbulence in this simulation; the only turbulence present is due to the streaming instability itself, which in turn produces $\alpha$-viscosity around $\alpha \approx 10^{-5}$. {\footnote{While that is inconsistent with our assumption that external turbulence is necessary to loft the silicate grains up in the disk atmosphere to remove the ice (see \sect{sect:limitations} on limitations of the model), the low turbulence only increases the number of small grains near the midplane, and is thus a conservative approach to what upper limit for the silicate fraction we should expect in the resulting planetesimals.}}  For Stokes number of $\St = 0.5$, this results in a dust scale height of $H_d \approx H/250$. We want to resolve this layer with at least 6 mesh points (the size of a stencil), and to this end, we have set the resolution of our simulation to be $N_x=N_y=N_z=256$ resulting in a points-per-scale-height value of $H/\Delta = 1280$.

\subsection{Simulation Parameters}
\label{sect:parameters}

We set the dust-to-gas mass ratio (metallicity) to be $Z = 0.03$, slightly above Solar, which is known to trigger strong clumping by the streaming instability \citep{Carrera+15, Yang+17, LiYoudin21}. The bulk mass of solids is evenly split into $N_p = 1.536 \times 10^7$ numerical super-particles; each particle's mass is the cumulative mass of the pebbles it represents, but the aerodynamical behavior of a particle is identical to that of a single pebble. We chose Stokes numbers such that ices and silicates are on cm and sub-mm size scales, respectively. Given

\begin{eqnarray}
    \St &\approx& 0.1 \left(\frac{a}{1 \rm \ cm}\right) \left(\frac{\rho_\bullet}{1 \rm \ g \ cm^{-3}}\right) \nonumber \left(\frac{T}{35 \rm \ K}\right)^{-1/2} \left(\frac{\rho_g}{10^{-13} \rm \ g \ cm^{-3}}\right)^{-1}\nonumber\\
&&
\end{eqnarray}

\noindent we assign half the particles a Stokes number of $\St = 0.5$, representing ices, and the other half a Stokes number of $\St = 5 \times 10^{-3}$, representing silicates.

For the pebble drift, we use the dimensionless parameter defined by \citet{BaiStone10b}

\begin{equation}
  \Pi\equiv \frac{\Delta\,v}{c_s},
\end{equation}

\noindent and set it to $\Pi=0.05$ (note also that $\Pi c_s = \eta v_K$). The relative strength of self-gravity to tidal shear is determined by another dimensionless parameter

\begin{equation}\label{eq:g_tilde}
    \widetilde{G} \equiv \frac{4\pi G \rho_0}{\varOmega_0^2},
\end{equation}

\noindent and is related to the usual Toomre $Q$ parameter \citep{Safronov60,Toomre64} by 

\begin{equation}\label{eq:toomre_q}
    Q = \sqrt{\frac{8}{\pi}}\widetilde{G}^{-1}.
\end{equation}

The constants on the right-hand side of \eq{eq:poisson} are non-dimensionalized and in code units are set to be $4\pi G = 0.1$, consequently setting the gravitational constant in code units to be $G=0.1 / 4 \pi$. Furthermore, with $\rho_0 = \varOmega_0 = 1$, then $\widetilde{G} = 0.1$ and $Q \approx 16$, justifying the exclusion of self-gravity from the equations of motion for gas. Gravitational collapse of the pebbles ensues once the pebble density within a mesh cell exceeds the Roche density,

\begin{equation}\label{eq:roche_density}
    \rho_{R} \equiv \frac{9 \varOmega_0^2}{4\pi G} = 9\rho_0 \widetilde{G}^{-1}.
\end{equation}

Once this happens, all the particles within a sphere of radius equal to one mesh cell are removed and replaced by a sink particle \citep{Johansen+15,Schafer+17} that has a Stokes number akin to that of a planetesimal and thus no longer feels gas drag. These sink particles can continue to accrete pebbles.

\subsection{The composition of icy pebbles}

We expect from sticking velocity arguments that the large pebbles will be mostly icy. This is because, starting from a bare silicate nucleus, and letting the grain aggregate icy and silicate monomers, a higher ice sticking velocity means that the grains would grow progressively icier as the silicate nucleus is diluted in the ice, forming ice-mantled silicate grains. If a 1\,mm bare silicate grain accretes only ice until cm-size, the volume ratio implies that the grain will be, per volume (mass), only 0.1\% (0.3\%) silicate. Allowing for a smaller bare silicate nucleus will increase the ice ratio, so we consider the bare silicate to only be a trace in the final composition. The main constraint will come from the concurrent coagulation of silicates and ices, which will depend on the coagulation efficiency \citep{LambrechtsJohansen14}.

The sticking velocity for water ice is expected to be about 10 times higher than silicates \citep{Wada+09,GundlachBlum15}. This conclusion was challenged by \citet{MusiolikWurm19}, who find that the high surface energy of ice grains applies only in a narrow temperature range near the ice line; below 175 K, the surface energy of water ice is akin to that of silicates. However, the translation from surface energy to sticking velocity is not straightforward, as collision outcomes also depend strongly on porosity, mass ratio, and impact parameter, with sticking collisions possible with velocities up to 100\,m/s \citep{Planes+20,Planes+21}. Also, \citet{Musiolik21} reports higher surface energies for irradiated ice due to the development of a liquid microfilm, depending on the width of the ice crust. We therefore consider the value suggested by \citet{GundlachBlum15}, 10\,m/s, for the sticking velocity for ice, and 1\,m/s for silicates. With 10 times higher efficiency of ice coagulation compared to silicate coagulation, we expect the matrix of the larger pebbles to be about 90\% ice by mass (silicates are denser than ice, by a factor 3, but ices are more abundant than silicates, by a factor 4, so the two effects almost neatly cancel).

\subsection{Pebble Accretion}\label{subsec:pebble_accretion}

\begin{figure*}
    \begin{center}
    \resizebox{\textwidth}{!}{\includegraphics{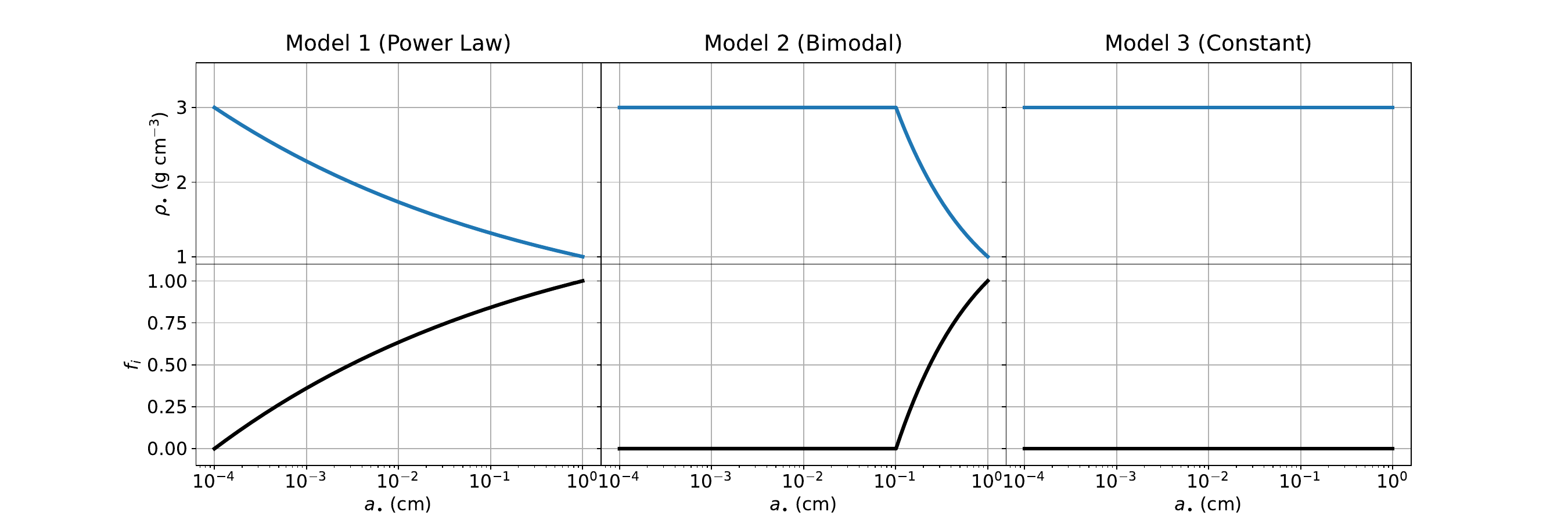}}
    \end{center}
    \caption{Internal density (upper panels) and ice volume fraction (lower panels) of accreted pebbles for the three models. (Left) The power law distribution in which the smallest grains are pure silicates and the largest pebbles are pure ice, with the trend that larger solids are icier. (Middle) The bimodal distribution in which pebbles between $\mu$m and mm are pure silicates, and pebbles above this size are increasingly icier with pebbles at one centimeter being pure ice. (Right) The constant distribution, in which pebbles of all grain sizes have a 0\% ice volume fraction and a density of $\rho_\bullet = 3$\,g\,cm$^{-3}$.}
\label{fig:dust_distribution}
\end{figure*}

\begin{figure*}
  \begin{center}
    \resizebox{\textwidth}{!}{\includegraphics{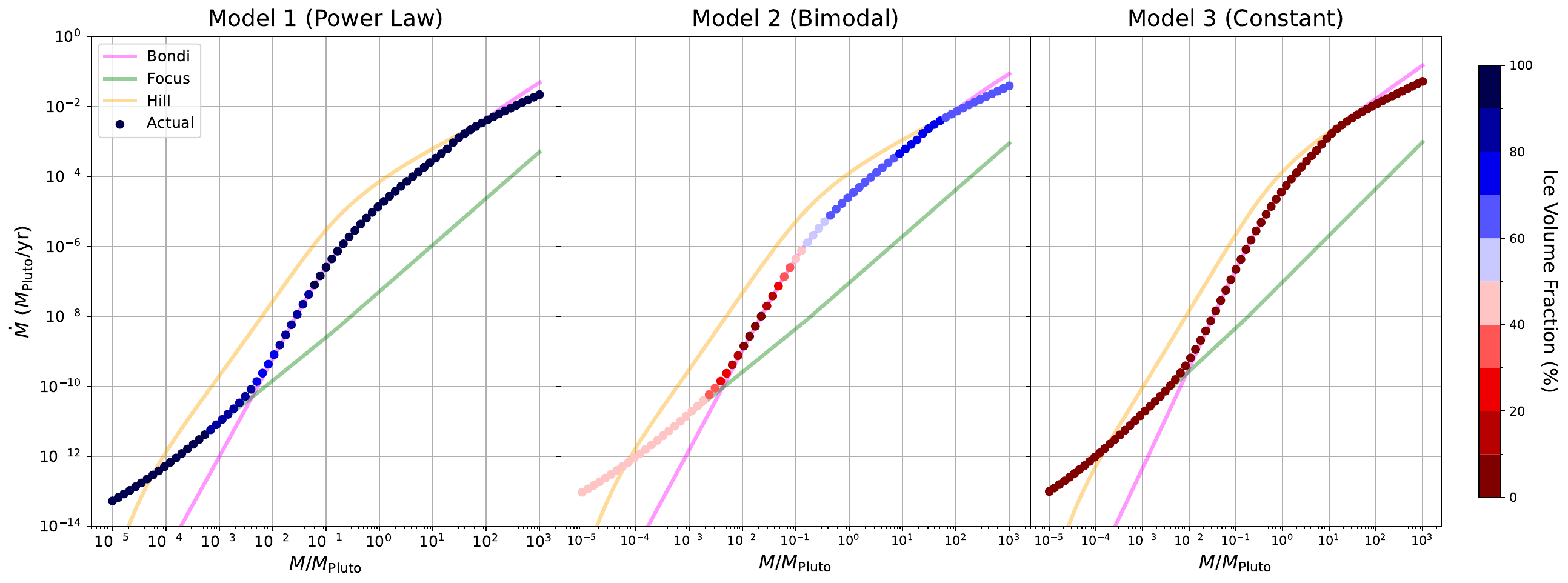}}
  \end{center}
\caption{The accretion rates for the three dust distributions at $t=0$. The orange lines are the accretion rates in the Hill regime, the magenta lines are the accretion rates in the Bondi regime, and the green lines are the accretion rates in the geometric/focusing regime. The scatter points are color-coded corresponding to the ice volume fraction of the pebble of most efficient accretion at the given protoplanet mass. (Left) Accretion rates of the power law distribution, which accrete no silicates except for a small window in the transition from geometric to Bondi, where roughly 20\% of the material being accreted is silicate. (Middle) The bimodal distribution begins accreting roughly 50\% ice and 50\% silicates in the geometric regime, but then accretes nearly 100\% silicates upon entering the Bondi regime. As a planetesimal grows through Bondi accretion, it begins to favor larger pebbles, thereby increasing the ice volume fraction that is being accreted, finally accreting mostly ices in the Hill regime. (Right) Accretion rates for the constant distribution, where all pebbles being accreted have a 0\% ice volume fraction.}
\label{fig:acc_rates}
\end{figure*}

Once gravitationally bound objects are produced, they continue to grow by accreting pebbles \citep{Lyra+08b,Lyra+09a,Lyra+09b,OrmelKlahr10,LambrechtsJohansen12}. Yet, for the objects produced, of the order of 100\,km in radius, the pebble accretion rates are much longer than we can model in a hydrodynamical simulation. Therefore, we take the planetesimal population produced in the streaming instability simulation, and feed them into a separate, relatively inexpensive, pebble accretion integrator, that solves the pebble accretion equations on evolutionary timescales.

The pebble accretion model we adopt considers a polydisperse distribution of pebbles, as recently developed by \cite{Lyra+23}, who found efficient pebble accretion on top of the direct products of streaming instability. Particularly, in the polydisperse model, the pebbles can have different internal density, which we will use as different composition. 

We consider three different models for pebble accretion, each differing in their respective internal density and ice volume fractions. The first model is a control, modeled as a power law

\beq
\rho_\bullet(a) = \rhoi \left( \frac{a}{a_{\rm max}} \right)^{-q_1},
\label{eq:powerlaw}
\eeq

\noindent with

\beq
q_1 \equiv \frac{\log (\rhos/\rhoi)}{\log (a_{\rm max}/ a_{\rm min})} > 0 
\label{eq:q1}
\eeq

\noindent so that the smallest grain has a density $\rhos$, and $\rho_{\bullet}$ decreases in internal density with increasing grain size such that the largest particle has a density $\rhoi$. We choose for this model $a_{\rm min} = 1\,\mu$m and $a_{\rm max} = 1\,$cm.

The second model is our physically motivated bimodal distribution

\begin{equation}
\rho_\bullet(a) =
    \begin{cases}
        \rhos, & \text{if } a < a_0\\
        \rhoi \left( \frac{a}{a_{\rm max}} \right)^{-q_2},& \text{if } a \geq a_0
      \end{cases}
\label{eq:secondmodel}
\end{equation}

\noindent with 

\beq
q_2 \equiv \frac{\log (\rhos/\rhoi)}{\log (a_{\rm max}/ a_0)} > 0.
\label{eq:q2}
\eeq

We choose for this model $a_0 =1$\,mm; that is, particles between the sizes $1\,\mu\text{m} \le a \le 1\,\text{ mm}$ are assigned internal density $\rhos$. Pebbles beyond 1\,mm in size then follow a power law such that the largest pebble has internal density $\rhoi$.

The last model assumes that all pebbles are silicates, with internal density $\rhos$. It  could represent a streaming instability filament of bare silicate pebbles and, as we will show, is necessary to reproduce Eris.
 
The models are illustrated in \fig{fig:dust_distribution}, where each column represents the different models. The upper row shows the internal density of the pebbles, and the bottom row is the ice volume fraction of the pebbles as a function of grain size. The ice volume fraction is calculated according to
\beqn
\rho = \rhoi f_i + \rhos f_s &&\label{eq:densityfractions}\\
f_i + f_s + f_v =1&& \label{eq:volumefractions}
\eeqn
where $f_j\equiv V_j/V$ is the volume fraction of component $j$; $V_j$ is the volume occupied by the component, and $V$ is the total volume. Here $f_v$ is the fractional volume of empty space, or ``porosity''. For compact pebbles, $f_v=0$ (and hence $\rho=\rho_\bullet$). Solving for the ice volume fraction of a pebble
  \beq
  f_i = 1-\left(\frac{\rho_\bullet-\rhoi}{\rhos-\rhoi}\right)
  \eeq
which is the quantity plotted in the bottom row of \fig{fig:dust_distribution}. We use $\rhoi=1$\,g\,cm$^{-3}$ and $\rhos=3$\,g\,cm$^{-3}$.

The accretion rates of each model at 20\,AU are illustrated in \fig{fig:acc_rates}, where the circles represent the actual accretion rates and they are color-coded according to the ice volume fraction of the pebbles that are most efficiently accreted at the given protoplanet mass. Initially the accretion rate is determined by the geometric cross section and gravitational focusing, with little contribution from gas drag (green line). When the Bondi radius becomes larger than the planetesimal, we enter the Bondi regime, where aerodynamic drag allows efficient capture of pebbles within the Bondi radius (magenta). Lastly, when the Hill radius becomes larger than the Bondi radius, we enter the Hill regime, in which the Hill radius becomes the new limiting accretion radius (orange). In the Bondi regime, the best accreted pebbles are those of stopping time similar to the time to cross the Bondi radius, which can be significantly smaller than the biggest pebbles present in the distribution \citep{Lyra+23}. This is of significant consequence because small bodies accrete in the Bondi regime \citep{Johansen+15}, opening the possibility of preferential accretion of small (silicate) pebbles for these bodies. Indeed, we see in \fig{fig:acc_rates} that the bimodal distribution (Model 2) shows a window of silicate accretion. Model 2 begins accreting pebbles of $\approx$ 50\%-50\% ices and silicates composition in the geometric regime, but then accretes nearly 100\% silicate pebbles right after the onset of Bondi accretion. This happens because the transition from the geometric regime to the Bondi regime is discontinous \citep{Ormel17}. In our model, the best accreted pebble at the geometric regime is of radius $\approx$3\,mm,  but it abruptly passes to $\approx$0.5\,mm after the onset of Bondi acccretion. The window of silicate accretion starts to narrow at higher protoplanet masses, that accrete larger pebbles, of higher ice volume fraction, finally accreting mostly ices in the Hill regime. Model 1 (power law) never accretes silicates significantly, as the pure silicate pebbles are too small. 

The mass accretion rates are integrated numerically choosing a timestep $\Delta t$ such that the mass accreted per timestep ($\dot{M} \Delta t$) is no greater than a fraction $C$ of the planetesimal mass $M_p$, 

\begin{equation}
    \Delta t = C \ \frac{M_p}{\dot{M}}.
\end{equation}

\noindent We set the value of $C$ at 0.01, found empirically to be a good compromise between stability and performance. At each time interval, we calculate the new mass and density of the planetesimal by taking a weighted average of the mass and density acquired by pebble accretion. Concurrently with pebble accretion, we reduce the gas and dust density exponentially in time, with an e-folding time of 2.5\,Myr. We terminate the simulation after 4 e-folding times (10 Myr), which we quote as ``the disk lifetime''. We note that at 5-10 Myr there is still some gas, yet much less than at $t=0$. The pebble accretion rates are impacted accordingly. To have significant mass accretion rates nearing 10\,Myr is unlikely in our model (because pebbles of all sizes are essentially decoupled), even though the calculations go until this time.

We highlight and address here the inconsistency of using a two-species pebble model for streaming instability, while using a continuous size distribution for the pebble accretion calculation. While inconsistent, this was done because our goal with the hydroydynamical simulation was to provide a proof-of-concept that the first planetesimals would be mostly icy, binning the pebbles in two species, one rocky and one icy, was judged a satisfactory first order approximation \citep[but see][for the impact of introducing multiple species in the development of the streaming instability]{Schaffer+18,Krapp+19,Paardekooper+20,Paardekooper+21,Schaffer+21,ZhuYang21,YangZhu21}. For the higher-mass objects, most of the mass is accreted during the pebble accretion phase.

\subsection{Porosity Evolution}

Finally, we must consider a compaction model to grasp the density evolution of Kuiper belt objects. We examine here the primary mechanisms by which planetesimals undergo removal of porosity. The first is compaction through gravitational pressure. As a planetesimal grows, gravity is continuously pulling all the material in the planetesimal toward the center. After some critical mass, around when the central pressure is greater than 10 MPa \citep{YasuiArakawa09,BiersonNimmo19}, the pull of gravity towards the center becomes strong enough so that the material yields, and compaction ensues, removing porosity.

The other mechanism involves the removal of porosity through heating. In the early solar nebula, there is a non-negligible amount of the short-lived radioactive isotope, $^{26}$Al \citep[initial abundance $^{26}$Al/$^{27}$Al = $\xtimes{5}{-5}$][]{MacPherson+95,Davidsson+16}. The decay of this isotope, if present in large quantities in the protoplanet, would act to essentially melt away porosity from these bodies. \citet{BiersonNimmo19} explore this effect in detail, along with gravitational compaction. We do not solve the full system of coupled non-linear partial differential equations; instead, we use the fact that regardless of the rock-to-ice mass fraction, \cite{BiersonNimmo19} find that the objects become fully compact at a radius of $R \approx 1500$\,km. Thus, we assume the following simplified porosity parametrization

\begin{equation}\label{eq:porosity}
  \phi(R) = \phi_0\ \theta (R_1-R) + \xi_{R_1,R_2}(R) \, \log_{10}\left(\frac{R_2}{R}\right) 
\end{equation}

\noindent where $R$ is the radius of the KBO, $R_2=1500$\,km, and $R_1=R_2/10^{\phi_0}$. Here $\theta$ is the Heaviside step function

\begin{equation}
\theta(x) \equiv \left\{
  \begin{array}{lr}
        1,  & \text{if } x > 0\\
        1/2,& \text{if } x = 0\\
        0,  & \text{if } x  < 0
  \end{array}
\right.
\label{eq:heaviside}
\end{equation}

\noindent and $\xi$ is the boxcar function

\beq
\xi_{x_1,x_2}(x) \equiv \theta(x-x_1) - \theta(x-x_2).
\label{eq:boxcar}
\eeq

\begin{figure*}
  \begin{center}
    \resizebox{\columnwidth}{!}{\includegraphics{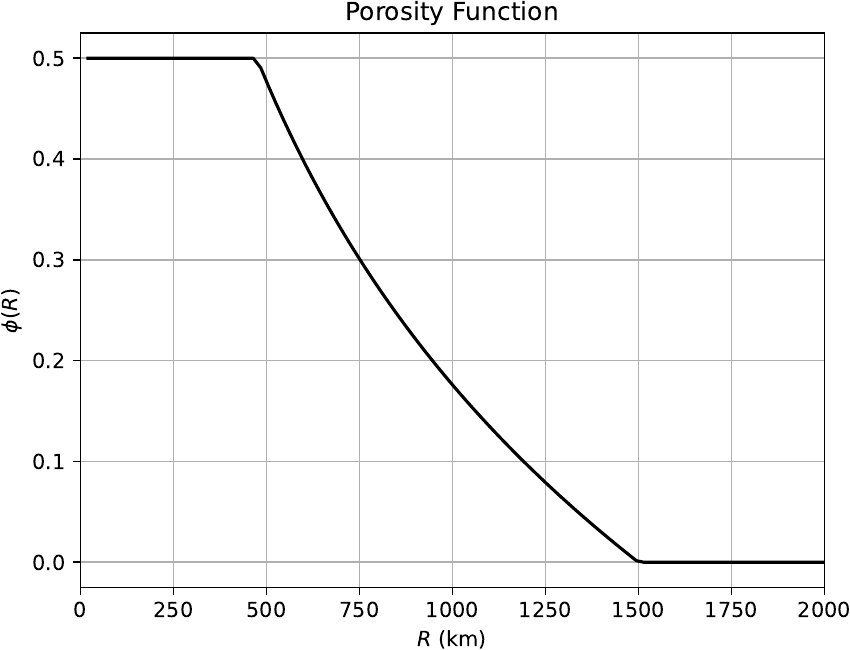}}
    \resizebox{\columnwidth}{!}{\includegraphics{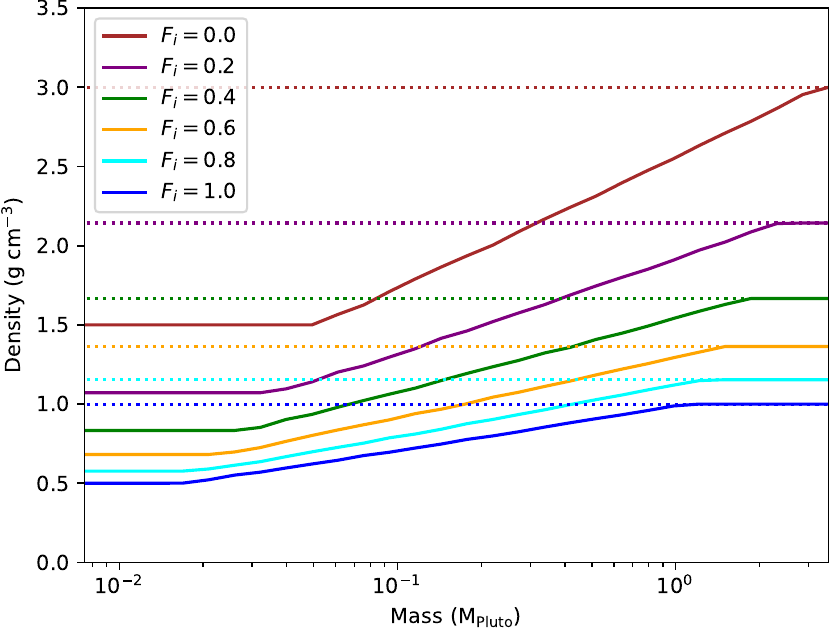}}
  \end{center}
\caption{Left: Graph of \eq{eq:porosity}, the parametrization we adopt for porosity. Right: Densities, for constant ice mass fraction $F_i$, calculated with \eq{eq:porosity} and \eq{eq:density-porosity} are shown in solid lines. The dashed lines of same color show the density of a fully compact body of same ice mass fraction.}
\label{fig:porosity_function}
\end{figure*}

\noindent \eq{eq:porosity} is plotted in \fig{fig:porosity_function}, left panel. It states that the porosity is constant at the porosity of formation $\phi_0$, up to a radius $R_1$, beyond which the porosity is removed, logarithimically, reaching zero at $R_2$. Choosing $R_2 = 1500$\,km and $\phi_0=0.5$ results in $R_1 \approx 474\,$km. Although admittedly crude, this approximation is sufficient for our purposes. We apply this model concurrently with pebble accretion, using the newly calculated mass and density to obtain a radius, and then using \eq{eq:porosity} to determine the porosity fraction. To aid the interpretation of our results, we first isolate the effect the porosity function would have on bodies of constant composition. The mass fraction of a constituent $j$ (ice or silicate) is defined as $F_j \equiv  m_j/M_p = f_j \rho_j / \rho $ where $m_j$ is the mass of the constituent in the body, $M_p=m_i+m_s$ is the total mass, and $f_j$ is the constituent's volume fraction. Given $f_v \equiv \phi$,  \eq{eq:volumefractions} can be solved for $\rho$ in terms of the ice mass fraction $F_i$ 
  
\begin{equation}
  \rho = \frac{\left(1-\phi\right)\rhoi\rhos}{\rhoi + F_i \left(\rhos-\rhoi\right)} 
  \label{eq:density-porosity}
\end{equation}

Curves of constant $F_i$ calculated according to \eq{eq:density-porosity} are shown in \fig{fig:porosity_function}, right panel. The full lines are porous bodies with porosity given by \eq{eq:porosity}; dashed horizontal lines mark fully compact (zero porosity) bodies.

\section{Results}
\label{sect:results}

\subsection{Planetesimal Formation}

We run a hydrodynamic simulation where we consider two pebble species, ices and silicates, with Stokes number $\rm{St} = 0.5$ and $\rm{St} = 5\times10^{-3}$, respectively. We run the simulation until planetesimal formation saturates (i.e., roughly an orbit goes by without pebbles collapsing into planetesimals). This condition occurs roughly after five orbits, or 1500 years, considering an orbit at 45\,AU is $T=2\pi/\varOmega_0 \approx 300\text{ years}$.

\begin{figure}
    \begin{center}
    \resizebox{\columnwidth}{!}{\includegraphics{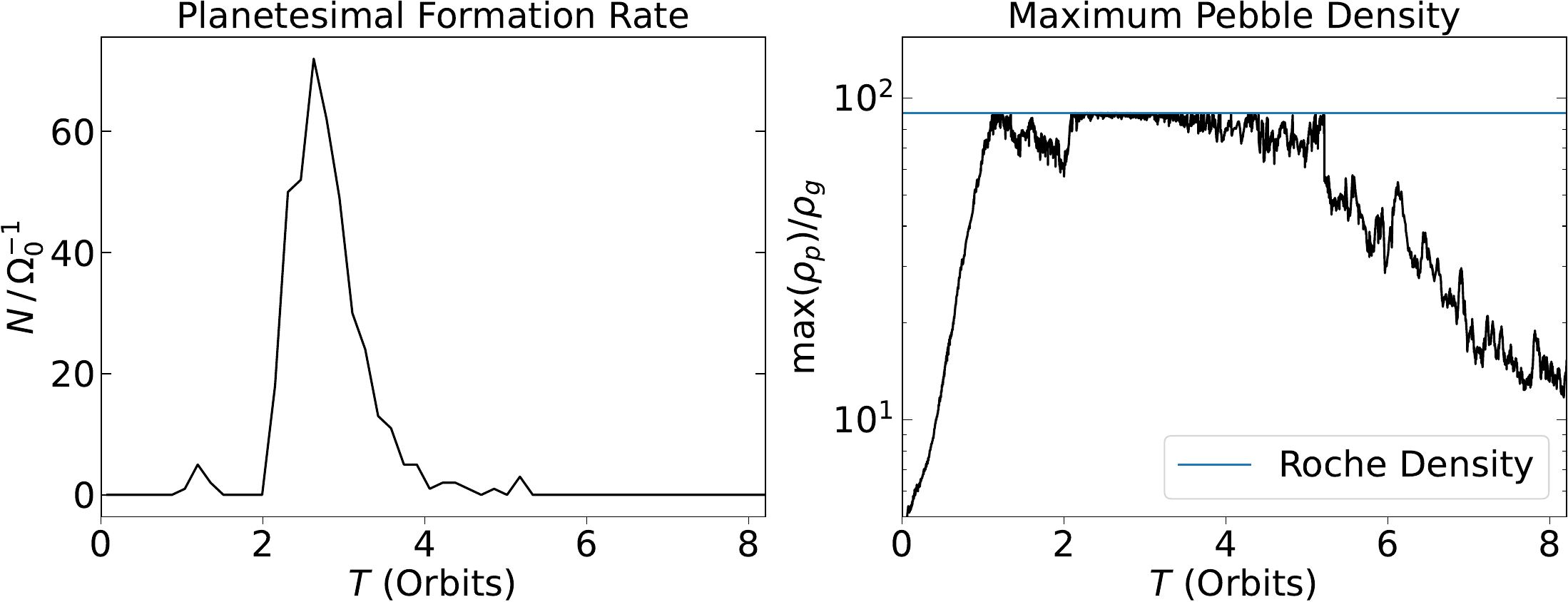}}
    \end{center}
    \caption{{\it Left:} Planetesimal formation rate (number of planetesimals $N$ formed per time interval $\varOmega_0^{-1}$), as a function of time, in the box. There is a rapid burst of planetesimal formation between two and four orbits and planetesimal formation comes to a halt after about five orbits. {\it Right:} Time evolution of maximum dust density in a mesh cell. The blue line corresponds to the Roche density, which if exceeded, particles in that mesh cell are removed and replaced by a sink particle. The initial rise in density from zero to one orbit marks the sedimentation of solids and concentration by the instability, after which the Roche density is achieved for several orbits but then drops after roughly five orbits.}
\label{fig:runA_rhopmax}
\end{figure}

This is visualized in \fig{fig:runA_rhopmax}, where the left panel shows the planetesimal formation rate as a function of time, and the right panel shows the maximum grain density within a mesh cell, also as a function of time. The left panel shows that between the second and fourth orbits, hundreds of planetesimals were formed, with a peak of 70 per time interval $\varOmega_0^{-1}$ ($\approx$ 50 yrs). The last planetesimal was formed right after 5 orbits (the small spike before the rate last goes to zero). The right panel shows that between the second and fourth orbit, there appears to be at least one mesh cell with pebble density achieving Roche density, almost continuously. After about five orbits, the maximum pebble density within a mesh cell steadily decreases, suggesting that pebble clumps will not reach the Roche density again. The density within a mesh cell is not allowed to exceed the Roche density, because once a mesh cell has achieved $\rho_R$, the particles within the accretion radius at formation (set to the mesh spacing $\Delta$) are removed and replaced by a sink particle. 

During the time elapsed, a total of 408 planetesimals are formed, 276 of which are accreted by other planetesimals, leaving 132 planetesimals by the end of the simulation. 

\begin{figure*}
    \begin{center}
      \resizebox{\textwidth}{!}{\includegraphics{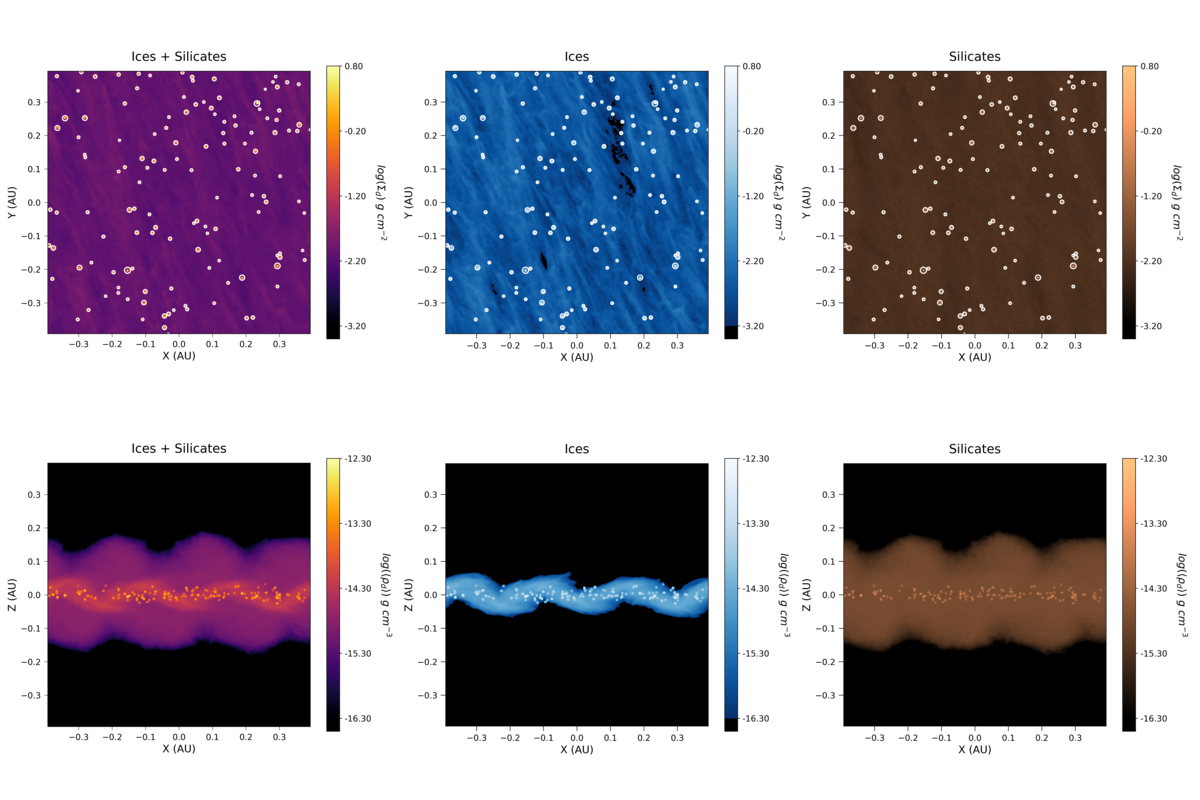}}      
    \end{center}
    \caption{The vertically integrated (top) and azimuthally integrated (bottom) dust density after six orbits, when planetesimal formation has saturated. The left panels correspond to the sum of ice and silicate densities, while the middle and right panels are the ice and silicate densities. Dots indicate formed planetestimals and circles in the top panels show the Hill radii of these planetesimals.  The Hill radii appear close to ice overdensities, consistent with our detailed measurements of high ice mass fractions.}
\label{fig:runA_rhop_ice_sil}
\end{figure*}

With icy pebbles being less susceptible to the drag force compared to silicates, they experience lower levels of turbulent diffusion, which provides support against stellar gravity. The result is that ices form a thinner, denser mid-plane layer compared to silicates \citep{Dubrulle+95,YoudinLithwick07} and are therefore more likely to form clumps that collapse into planetesimals. This is better demonstrated in \fig{fig:runA_rhop_ice_sil}, where the top row of panels shows the integrated column density along the z-axis, and the bottom row the volume density averaged over the y-axis. The first column in each row shows the combined density of ices and silicates, while in the second and third columns we disaggregate the pebbles into ices and silicates, respectively. The circled objects are planetesimals and the size of the circles represents the Hill radius of the encapsulated planetesimal.

We see in the column density plots that the filamentary structure associated with the streaming instability is apparent only in the ice plots, whereas the silicate plot shows a smoother distribution. This is because the silicates are too tightly coupled to the gas and do not drift as much as the ices, being thus less prone to the streaming instability \citep[e.g.][]{YangZhu21}. The azimuthal average plots show that the silicates have a height of $H_d \approx 0.1$\,AU (true bare silicates in stronger turbulence would have a taller scale height), whereas ices have a denser layer that is more favorable to the formation of planetesimals.

To help distinguish between these two processes (vertical settling vs streaming), we analyze the time-evolution of the ice-to-silicate ratio. The settling time for a grain of a given Stokes number is \citep{Youdin10}

\beq
t_{\rm settle} = \frac{1+2\St^2}{\varOmega\St},
\label{eq:settling}
\eeq

\begin{figure}
    \begin{center}
    \resizebox{\columnwidth}{!}{\includegraphics{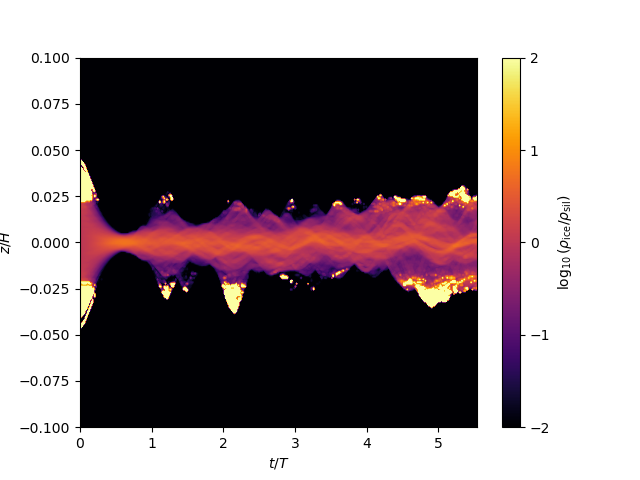}}
    \end{center}
    \caption{Evolution of the ice-to-silicate ratio in the vertical plane. The bright yellow is saturated. The ices settle in about 0.5 orbit, reaching an ice-to-silicate ratio of about 10 in the midplane. The filamentary structure of the streaming instability is seen after $\sim$1 orbit.}
\label{fig:icesil_zplane}
\end{figure}

\begin{figure*}
    \begin{center}
      \resizebox{\columnwidth}{!}{\includegraphics{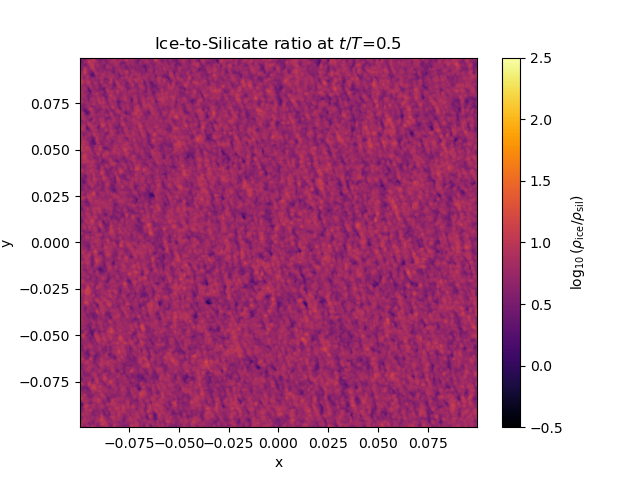}}
      \resizebox{\columnwidth}{!}{\includegraphics{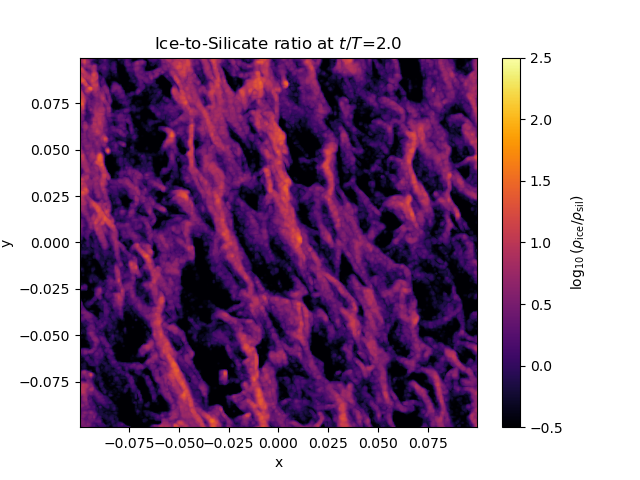}}
    \end{center}
    \caption{{\it Left:} Ice-to-silicate ratio in the midplane at $t = 0.5$ orbits, approximately the time when the ices settle. {\it Right:} Ice-to-silicate ratio in the midplane at $t = 2.0$ orbits, approximately the time of formation of the first planetesimals. The increase in ice-to-silicate due to settling leads to a factor 10 enhancement. Further enhancement is due to the streaming instability.}
\label{fig:icesil_midplane}
\end{figure*}

\noindent which yields $\sim$0.3 and 30 orbits, for $\St=0.5$ and $\St=5\times 10^{-3}$, respectively. 
We started the simulation with both dust species at Gaussian stratifications of 0.01H, which is the equilibrium scale height for the silicate dust grains for $\alpha\sim 10^{-6}$. As a result, the silicate grains do not evolve significantly vertically. The ice pebbles settle very fast. We show in \fig{fig:icesil_zplane} the time evolution of the ice-to-silicate ratio (defined as $\rho_{\rm ice}/\rho_{\rm sil}$, where $\rho_{\rm ice}$ and $\rho_{\rm sil}$ are the volume densities of ices and silicates, respectively) in the vertical plane. It mimics the evolution of the ice pebbles. Indeed we see that the settling time is $\sim$0.5 orbit, as expected. At 1--2 orbits, the streaming instability develops and saturates.

We show in \fig{fig:icesil_midplane} the midplane snapshots of the ice/silicate ratio at $t=0.5$ and 2 orbits. These are the times at $t_{\rm settle}$ for the ices, and when the first planetesimal form. The left panel shows the ratio resulting from settling alone. As we see it, it looks homogeneous, with ice-to-silicate ratio about 10. The right panel shows the filamentary structure expected from streaming instability, showing the difference in ice-to-silicate ratio between the filaments and the voids.

\begin{figure}
    \begin{center}
    \resizebox{\columnwidth}{!}{\includegraphics{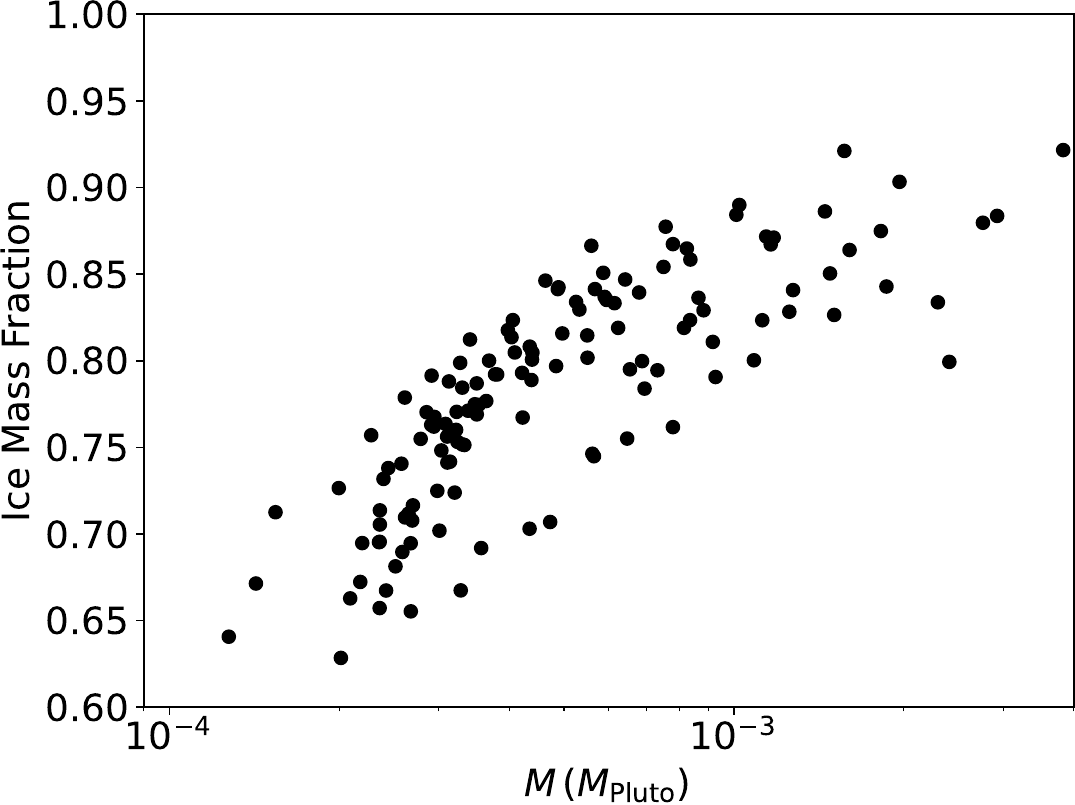}}
    \end{center}
    \caption{Ice mass fraction of the planetesimals formed at the end of the streaming instability simulations. The composition is not uniform; a trend emerges where the higher mass planetesimals are formed with higher ice mass fraction.}
\label{fig:mass-vs-ice}
\end{figure}

Finally, we show in \fig{fig:mass-vs-ice} the mass and ice mass fraction $F_i$ of the planetesimals, at the end of the simulation. The masses range from $\approx\xtimes{1.5}{-4}$ to $\approx\xtimes{3}{-3}$ \mplutoc, with a trend that the lower mass planetesimals are more silicate-rich, and the larger ones more ice-rich. The trend does not seem to be a numerical artifact, a point we will go back to in \sect{sect:disc-planetesimals}. We estimate in \app{app:heating} whether this ice mass fraction would lead to melting from heating from $^{\rm 26}$Al.

\subsection{Integrating Pebble Accretion}
\label{sect:results-pebble}

We take the distribution of planetesimals we found in the previous section, and feed it into a polydisperse pebble accretion integrator \citep{Lyra+23}. The streaming instability model was calculated at 45\,AU, yet at that location the pebble accretion rates are too slow, and we do not expect planetesimals formed at that distance to become protoplanets, given the existence of the ``cold classical'' KBOs population \citep{Brown01,Kavelaars+08,Petit+11}. In fact, in the context of the ``Nice'' model \citep[][see \sect{subsec:nice_model}]{Tsiganis+05,Morbidelli+05,Gomes+05} Neptune's current location at 30\,AU constrains that all KBOs aside from the cold classicals formed up to 30\,AU, otherwise Neptune would have continued its migration further out \citep[][and references therein]{Stern+18}. Therefore, we vary the distance at which we perform the pebble accretion integration, from 10 to 30\,AU, in intervals of 5\,AU. We also complete the initial mass function of planetesimals up to $10^{-2}$ \mpluto (following the composition trend found in the ices and silicates streaming intability model), given that this is the approximate mass for the onset of pebble accretion in the Bondi regime \citep{Lyra+23}.

\begin{figure*}
    \begin{center}
      \resizebox{\textwidth}{!}{\includegraphics{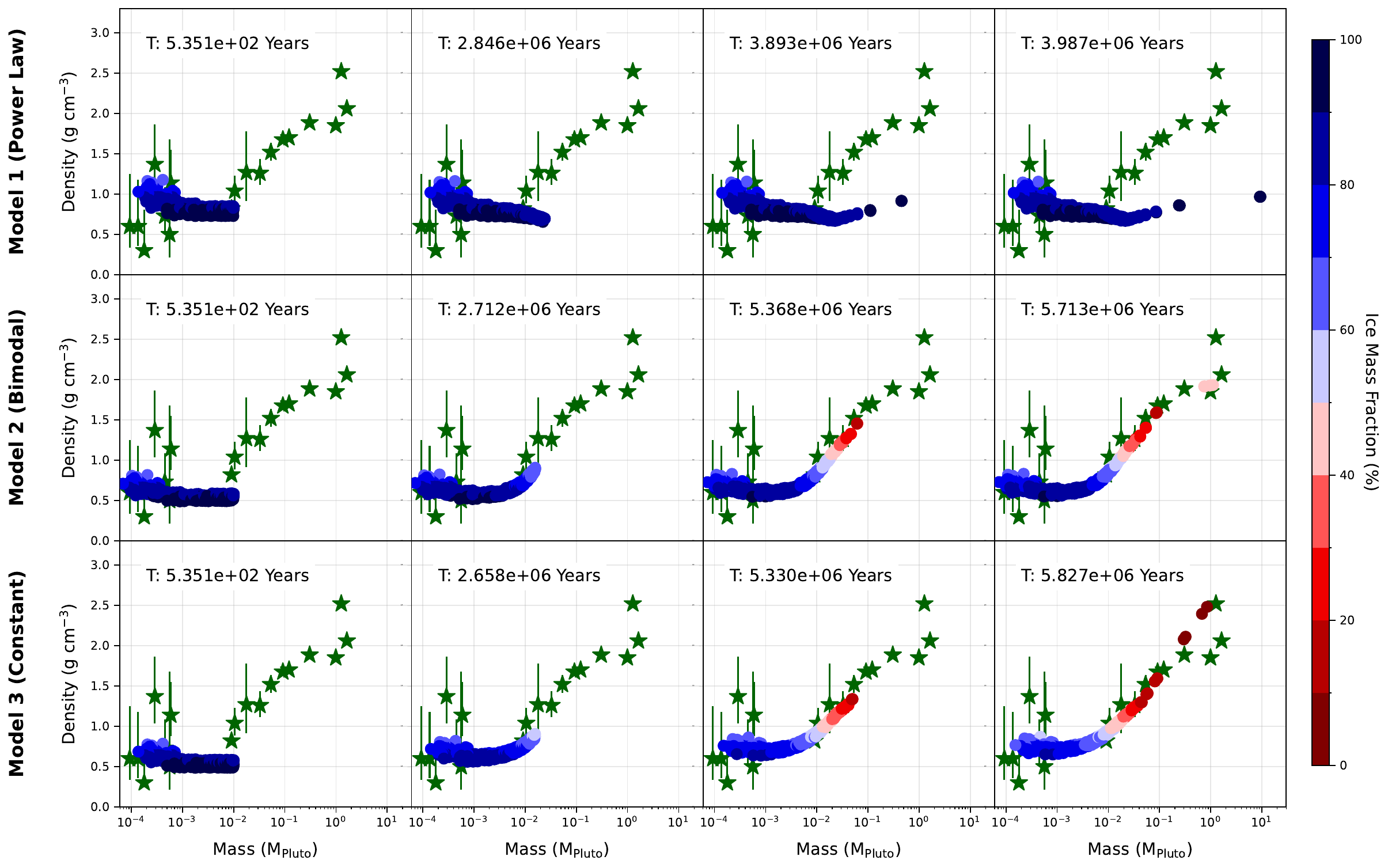}}
  \end{center}
  \caption{{\it Upper row:} Mass, density, and ice mass fraction evolution of protoplanets with pebble accretion of the power-law density distribution (model 1, \eqs{eq:powerlaw}{eq:q1}). Actual KBO data is overplotted as green stars. The best accreted pebbles at each mass are still too icy to lead to significant density increase through silicate accretion. The most massive planets formed within 5\,Myr are of Pluto mass, but their composition is almost pure water ice.\\
   {\it Middle row:} Same as upper row, but for the bimodal density distribution (model 2, \eqs{eq:secondmodel}{eq:q2}). As the intermmediate-mass objects ($10^{-2}-10^{-1}$ \mplutoc) accrete silicate pebbles, the mass-density trend is reproduced. The flattening at higher mass is also reproduced, matching Haumea, Pluto, and Triton. The model does not reproduce the density of Eris.\\
  {\it Lower row:} Same as the other rows, but for constant (silicate) pebble density (model 3). The intermmediate-mass objects ($10^{-2}-10^{-1}$ \mplutoc) is similar to model 2, but the model overshoots the density of the higher mass objects Haumea, Pluto, and Triton. Eris, however, is reproduced.}
\label{fig:allmodelsresults}
\end{figure*}

We show in the upper row of \fig{fig:allmodelsresults} the result for the power law distribution (model 1), at 20\,AU. The panels are a time series, showing the mass vs density evolution, and color-coded the ice mass fraction of the protoplanets. The actual KBO data from \fig{fig:kbos_w_label} are overplotted as green stars. We see that this model favors ices too strongly to cause any significant changes in density, even with the removal of porosity through compaction. The silicate pebbles in this model are simply too small to be accreted efficiently. Pluto mass is achieved around 4\,Myr, but consisting of almost completely pure water ice. 

The results change considerably when using the bimodal density distribution (model 2, middle row of \fig{fig:allmodelsresults}). In the figure, we see that the growth rate matches the mass-density trend, providing an excellent fit to the intermediate-mass range objects from $10^{-2}$ (2002 UX$_{\rm 25}$) to $10^{-1}$ \mpluto (Charon), and also the higher-mass objects, Haumea, Pluto, and Triton. Yet, the model fails to reproduce the larger density of Eris. 

Motivated to reach the density of Eris, we try a 3rd model, which consists of accretion of pure silicate pebbles. The results are shown in the lower panels of \fig{fig:allmodelsresults}. Starting from the icy planetesimals produced by the streaming instability, the density increases monotonically with mass as only silicates are accreted. This model also provides a good fit to the observed mass-density trend for intermediate-mass KBOs, up to $10^{-1}$ \mpluto (evidencing that model 2 was accreting silicates in this mass range, as we will expand on in \sect{sect:silicatewindown}). However, beyond this threshold, unlike the bimodal distribution, the objects have an ice mass fraction that is close to zero. The model largely overestimates the densities of Haumea, Pluto, and Triton. Yet, it does reproduce the high density of Eris. 

\subsubsection{The effect of distance}

We explore now the parameter space of distance in the pebble accretion model, taking Model 2 as fiducial and varying radii from 10\,AU to 30\,AU, at 5\,AU intervals. We illustrate the results in \fig{fig:kbo_v_r}. This figure shows that we are able to best reproduce the mass ranges of KBOs if they formed between 15\,AU and 22\,AU. At 10\,AU, we see that within 500 thousand years, masses attained through pebble accretion are a factor of 10 larger than the mass of Pluto. Considering the lifetime of the solar nebula is of order 10 million years (i.e., a few e-folding times), it is expected that the masses at this heliocentric distance will ultimately be several orders of magnitude larger than the mass of Pluto. On the other hand, considering the accretion rates at 30\,AU, we see little to no growth from pebble accretion. After 10 million years, planetesimals barely grow to $0.2\,M_{\rm{Pluto}}$, suggesting that it is unlikely that Pluto and the rest of the dwarf planets formed at or beyond 30\,AU. A favorable location is between 15\,AU and 22\,AU, where Pluto's mass is reached within 10 million years. We conclude that for planetesimals formed beyond this range pebble accretion rates would be far too low to achieve Pluto mass within the assumed lifetime of the solar nebula (as also found by \citealt{Lambrechts+14} for monodisperse pebble accretion). Conversely, if the planetesimals formed any closer to the Sun, then high accretion rates will produce planetesimals much larger than Pluto, contrasting the masses seen in the Kuiper belt.

\section{Discussion}
\label{sect:discussion}

\subsection{The Importance of the Silicate Window}
\label{sect:silicatewindown}

\begin{figure*}
      \begin{center}
    \resizebox{\textwidth}{!}{\includegraphics{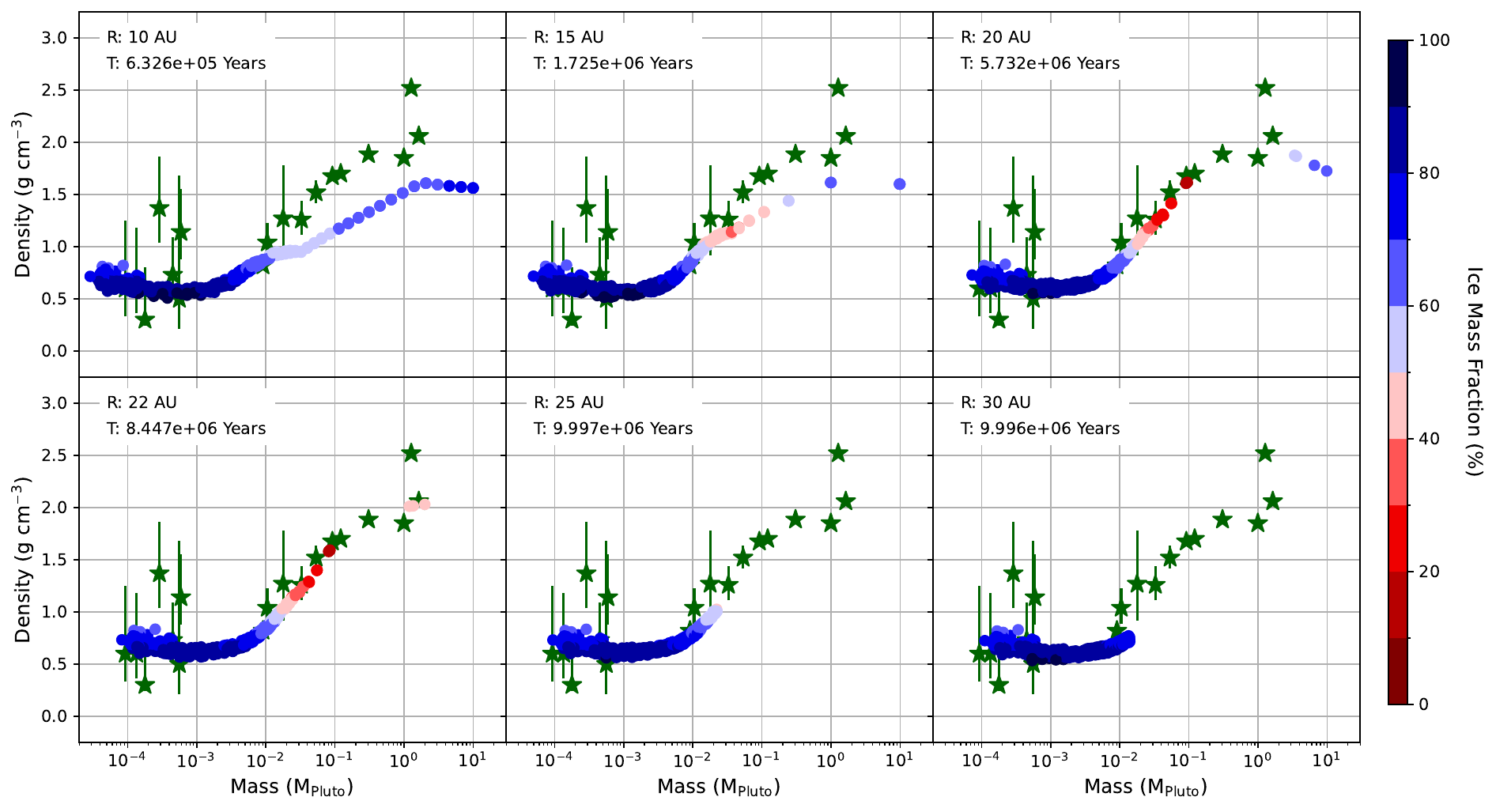}}
    \end{center}
\caption{The results from applying the polydisperse pebble accretion model with the bimodal distribution of pebbles across various heliocentric distances. At 10\,AU (upper left) not only do we underestimate the densities of the KBOs, but the masses produced are much larger than those seen in the Kuiper belt. At just 500 thousand years, planetesimals are already an order of magnitude larger than Pluto. From 15\,AU to 22\,AU, we fit the density trend well and reach Pluto mass in 1.8 million years, 5.7 million years, and 8.5 million years, respectively. Beyond 22\,AU, accretion rates are so low that we can not reach Pluto's mass within the assumed lifetime of the disk (10 million years).} 
\label{fig:kbo_v_r}
\end{figure*}

\begin{figure*}
  \begin{center}
    \resizebox{\textwidth}{!}{\includegraphics{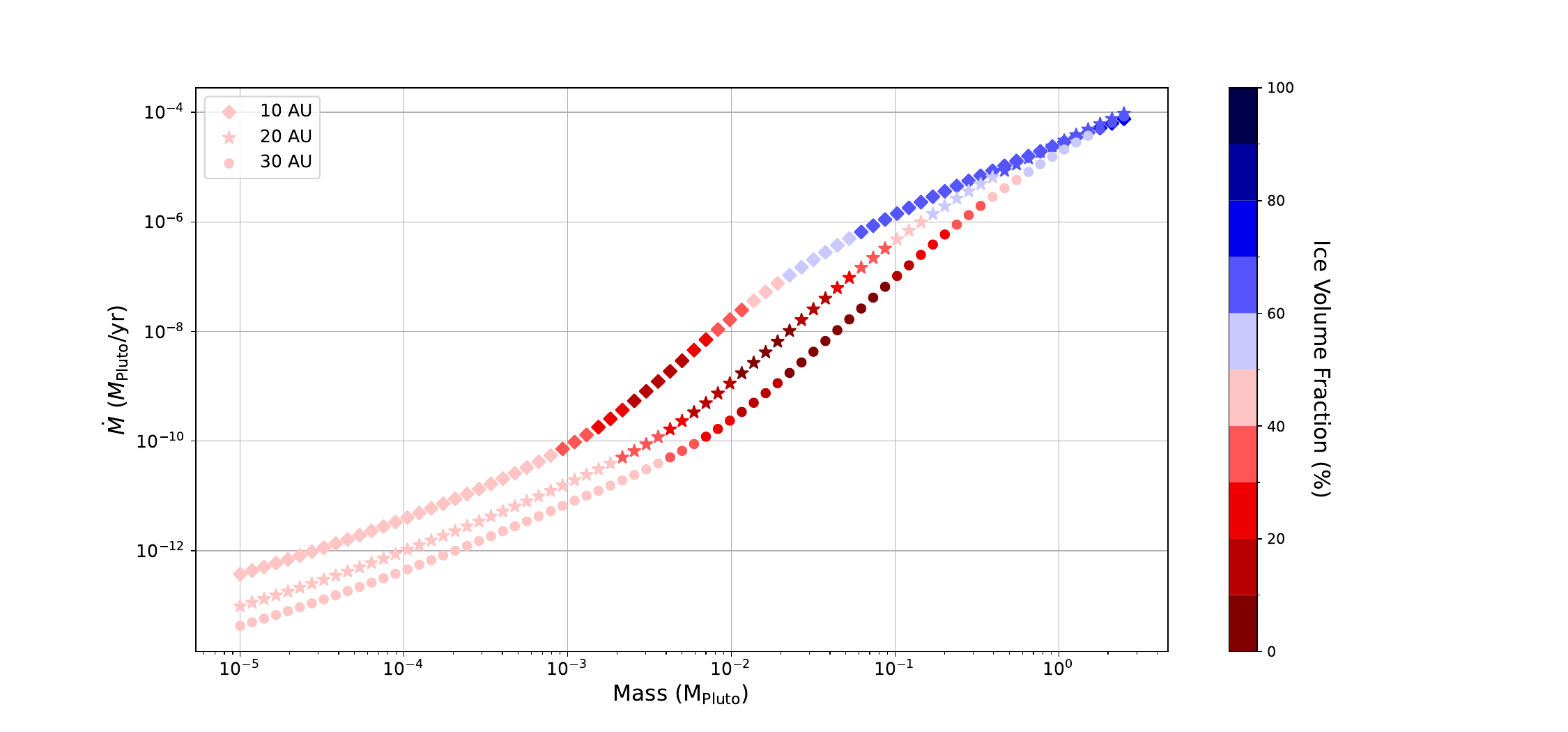}}
  \end{center}
\caption{Comparison of accretion rates for the bimodal distribution at different radii. The window at which silicate accretion is most effective shifts to higher masses as we move outwards in the disk. This provides further evidence for 20\,AU being a favorable location, as silicate accretion is most effective around $10^{-2} M_{\rm Pluto}$, right when the densities of Kuiper belt objects begin to increase.}
\label{fig:acc_v_r}
\end{figure*}

To understand these results, we show in \fig{fig:acc_v_r} the comparison of the accretion rates at 10 (diamonds), 20 (stars), and 30 (circles)\,AU, for the bimodal distribution. We see that at $\sim 10^{-2}$ \mplutoc, roughly the mass where Kuiper belt objects begin to show an increase in density, the accretion rate at 10\,AU is roughly an order of magnitude larger than the accretion rate at 20\,AU, and about two orders of magnitude larger than the accretion rate at 30\,AU. Furthermore, we find that the window of enhanced silicate accretion provided by the Bondi regime (see \sect{subsec:pebble_accretion}) shifts to higher masses with increasing orbital distance. The result is that at 10\,AU, planetesimals with mass $M=10^{-2}M_{\rm{Pluto}}$ have already missed the window of silicate accretion, resulting in the lower densities seen in the left-most panel of Fig. \ref{fig:kbo_v_r}. Conversely, at 30\,AU, the window of silicate accretion extends beyond $10^{-1}$ \mplutoc, which could facilitate this model's ability to reproduce the density of Eris, but due to the low accretion rates, these masses are never achieved.

The figure also shows why both model 2 and model 3 reproduce the mass-density trend for intermmediate masses. This is because at $10^{-2}$ \mpluto at 20\,AU, model 2 is accreting almost pure silicates, like model 3. At $10^{-1}$ \mpluto the models diverge significantly as the larger pebbles are icy in model 2.

\subsection{The planetesimals formed}
\label{sect:disc-planetesimals}

We perform a streaming instability simulation at 45\,AU (a location where no significant pebble accretion is expected), producing the population of planetesimals seen in \fig{fig:mass-vs-ice}. The span in mass is about an order of magnitude, from $\approx\xtimes{1.5}{-4}$ to $\approx\xtimes{3}{-3}$ \mplutoc.  The trend seen in that figure is that smaller planetesimals are more silicate-rich than the larger ones. This is not a numerical artifact, as Bondi accretion is too slow at this mass to significantly affect the masses over the timespan of the hydrodynamical simulation. Also, the Hill radii of these planetesimals are resolved. For 45\,AU and at this resolution, the mass at which the Hill radius $R_H$ is equal to the length $\Delta$ of a mesh cell is

\beqn
M_p^{\left(R_H = \Delta\right)}  &=& 3\,M_\sun \left(\frac{\Delta}{r}\right)^3\\\nonumber
&\approx& \xtimes{8.6}{-5}\,M_{\rm Pluto}
\eeqn

\noindent or about half the mass of the smallest planetesimal formed. The trend seen in \fig{fig:mass-vs-ice} is likely physical. We note that this is consistent with the fact that some small objects are high in density, such as (66652) Borasisi \citep[diameter $\approx$ 160\,km, mean density $\approx$ 2.1\,g\,cm$^{-3}$][]{Trujillo+00,Noll+04}. The effect of radiogenic heating is a significant function of ice mass fraction, because $^{26}$Al is only present in dust. Naively, we would expect that the smaller objects would be less affected by radiogenic heating (bulk heating versus area cooling results in $\propto R$ dependency), but \cite{Davidsson+16} note that small objects have poor thermal conduction than larger objects, trapping heat and melting the interior if the ice mass fraction is low. The melting removes the porosity, making these objects high density. In our model, significant dispersion in ice mass fraction exists for streaming instability products less massive than $10^{-3} M_{\rm Pluto}$, so some objects (the more ice-rich) should avoid melting whereas some (the more rock-rich) should undergo melting. The scatter for the low mass objects is consistent with \fig{fig:kbos_w_label}. All streaming instability products more massive than $10^{-3} M_{\rm Pluto}$ are ice-rich and should avoid melting.

\subsection{CO ice line}

We take this planetesimal population, and feed them into a polydisperse pebble accretion calculation, at different locations, using three different models for a pebble size-composition relation, as shown in \fig{fig:allmodelsresults}. We find that the density and masses of large KBOs are best reproduced if we place their formation between 15 and 22\,AU. Interestingly enough, this distance roughly coincides with the probable location, at 20\,AU, of the CO snowline. 

A best location scenario coinciding with the general location of the CO iceline opens the possibility that CO ice could also be relevant for the composition of the pebbles and hence, the densities of KBOs. While CO is a hypervolatile, CO reacts with water to produce methanol \citep{Grundy+20}, which is refractory. Having a density of $0.64$ g\,cm$^{-3}$ at 20 K, methanol ice could also explain the low densities of the small KBOs, if a significant fraction of their bulk is of that material. Indeed, near an ice line pebble growth is boosted through deposition and nucleation of vapor onto bare silicate and already ice-covered pebbles \citep[see ][ in the context of water ice]{Ros+19}. This possibility is consistent with the lack of observed water ice spectral features on small KBOs \citep{Barucci+11,Grundy+20}.

The CO ice line is also favorable to the formation of the ice giants, not only owing to the excess of carbon and lack of nitrogen found in the ice giants \citep{Ali-Dib+14} but also because ice giants are thought to contain $> 10\%$ methane by mass, which would occur if the ice giants formed between the CO and N$_2$ ice lines \citep{Dodson-RobinsonBodenheimer10}. Our results are also in excellent agreement with the constrains that Pluto and the KBOs (except the cold classicals) formed closer in, as previously discussed in \sect{sect:results-pebble} \citep[see also][]{Malhotra93,Stern+18,Canup+20}. 

\subsection{No special timing}
\label{sect:special-timing}

Our model also has the advantage of not needing to invoke a special formation time for the Kuiper belt objects, contrasting with the model from \citet{BiersonNimmo19}. This is because silicates, or more specifically the short-lived radioactive isotope $^{26}$Al with a half-life of $7\times10^5$~yr, are not incorporated in the initial formation of KBOs. Instead, they are incorporated through Bondi accretion in the later stages of the evolution of KBOs, when a large fraction of $^{26}$Al has already been depleted. Lastly, since we were able to reproduce the high density of Eris through the pebble distribution model of pure silicates, our model could suggest that Eris formed from a streaming instability filament \citep{LorekJohansen22} with a large fraction of rock, or that Eris formed at a later time when volatiles in the disk were lost through drift and photoevaporation (although that would potentially not leave enouh time for pebble accretion to operate). Alternatively, Eris could have formed with the density predicted by model 2, and subsequently lost some of its ice mantle through collisional evolution \citep{BarrSchwamb16}. 

\subsection{Connection to the Nice Model}\label{subsec:nice_model}

Our model is in agreement with the ``Nice'' model scenario \citep{Tsiganis+05, Morbidelli+05, Gomes+05, Emel'yanenko22}. In this picture, after the dissipation of the solar nebula, the giant planets are initially in a more compact configuration than currently, and their masses decreasing monotonically as heliocentric distance increases (i.e., Uranus and Neptune swapped). Jupiter is placed in the vicinity of its current location at 5\,AU, Uranus around 11 - 17\,AU, and Saturn and Neptune between these two limits. The giant planets are also assumed to have been in near circular and co-planar orbits.

A belt of protoplanets exists just beyond the orbit of the outermost giant planet (in this case Uranus), and objects in the inner edge of the belt interact with the planet, being scattered inward or outward, exchanging angular momentum \citep{FernandezIp84}. The net gain of angular momentum of Uranus would be zero in this scenario, but the presence of Jupiter breaks the symmetry. A protoplanet scattered outward will likely return to interact with the planet again, but a protoplanet scattered inward has a high chance of interacting with Jupiter and getting ejected out of the solar system. The inner disk is thus a better sink of angular momentum than the outer disk, and the net result is that Uranus migrates outward, whereas Jupiter migrates inward. Adding Neptune and Saturn does not alter the conclusion; these planets also migrate outward as they scatter protoplanets toward Jupiter, than in turn ejects them. Once Jupiter and Saturn cross their 2:1 mean orbital resonance, dynamical instability ensues, throwing the ice giants into the primordial belt, and implanting a small fraction (0.01\%-0.1\%) of the objects into the present-day Kuiper belt (except the cold-classicals, that likely formed in situ). Further interaction with the belt dynamically cooled the orbits of the giant planets post-instability. The original belt extended at most up to 30\,AU, the current orbit of Neptune (otherwise Neptune would have migrated further outward). 

Our results are in excellent agreement with this model. While planetesimals can form by streaming instability at large heliocentric distances, explaining the cold classicals, pebble accretion is only efficient up to $\approx$25\,AU \citep{Johansen+15}. Beyond this distances, planetesimals do not grow up to Triton or Pluto sizes, even by polydisperse pebble accretion \cite{Lyra+23}, as we have demonstrated. Scattering small planetesimals is not efficient to drive further migration, so Neptune's current location coincides with where pebble accretion stops being efficient at growing large objects. 

\section{Limitations and Future Work}
\label{sect:limitations}

The model presented is an exploratory proof-of-concept model, and as such has a number of simplications.

First and foremost, we argued in the introduction that compositional differences in the pebbles should be expected, because of photodesorption of ices off small dust grains. The compositional difference can only be maintained if the photodesorption rate and coagulation rate are similar, which a growth time estimate indeed shows they are. The growth time of dust grains, up to a factor of order unity,  is $t_{\rm grow} \sim \varepsilon^{-1} T_{\rm orb}$ \citep{Birnstiel+12a,LambrechtsJohansen14,LorekJohansen22}, where $T_{\rm orb}$ is the orbital period. The disk starts with $\varepsilon\sim 10^{-2}$ of dust, of interstellar (sub-micron) size. Assuming an orbital period $T_{\rm orb}\sim 100$ years yields $t_{\rm grow}\sim 10^4$ years. The dust thus coagulates very quickly into pebbles. The steady state, however, is top-heavy, leaving about $\varepsilon\sim 10^{-4}$ of sub-micron dust, most of the mass residing in larger pebbles. Under these conditions, after most small grains are consumed, the growth time then jumps to $\sim$1 Myr, which is similar to the photodesorption rate.

Despite this assurance, the argument remains mostly qualitative, and thus the conversion between radius and composition we used is arbitrary. A future model should include coagulation, drift, fragmentation, turbulence, UV and cosmic ray desorption, calculating the grain size vs composition relation from first principles. Such a model would also have the benefit of constraining the transition in the bimodal model, which we arbitrarily placed at 1\,mm.

Secondly, it is an inconsistency that we used a two-species pebble model for the streaming instability, whereas the pebble accretion integrator used a continuum distribution. Ideally, the two calculations should be consistent, with the streaming instabiliy model also using a continuum of pebble sizes, of mixed composition. The computational cost of three-dimensional hydrodynamical streaming instability simulations also motivated this simplification.

Since we found that the best location for formation is near the CO ice line, our pebble accretion model could benefit from evolving the CO abundance, matching chemical evolutionary expectations, and the depletion of volatiles from heating, radial drift, and disk winds.

It is also an idealization that our model assumes an infinite supply of pebbles whereas, realistically, the pebbles drift and are eventually lost if there is no re-supply. Interestingly, this would perhaps lead to a compositional variation in time, because the larger pebble drift faster than the smaller ones. At earlier times, when planetesimals first form, ice allows for the formation of large pebbles that drift, rapidly depleting the ice. Pebble accretion, conversely, is a slower process, so it would mostly occur after ice depletion in the outer disk, feeding from the remaining smaller, more silicate-rich, dust grains. Pebble drift would be specially important for the evolution of CO-coated pebbles, because once they drift inward to the CO ice line, the CO gas evaporates and is vulnerable to disk winds, photoevaporation, or photodissociation.

Our model would also benefit from a more involved porosity evolution model that takes into consideration the abundance of radioactive elements like $^{26}$Al and the impact that radioactive heating has on the porosity of KBOs.

\section{Conclusions}
\label{sect:conclusions}

We set out to reproduce the density trend of Kuiper belt objects, where small objects have densities less than water ice and large Kuiper belt objects can reach densities of $2.5 \text{ g cm}^{-3}$. We run a hydrodynamic simulation to model planetesimal formation via the streaming instability of ices and silicates, where silicates are small grains with short friction times, and ices are large grains with long friction times. We find that planetesimals formed under these conditions are icy and are low in mass $M_p \lesssim 10^{-2} M_{\text{Pluto}}$, effectively reproducing the densities observed in the low mass Kuiper belt objects. We also find a correlation between ice mass fraction and planetesimal mass, albeit with significant dispersion. 

We then use these planetesimals formed by the streaming instability as starting masses for the recently derived polydisperse pebble accretion model \citep{Lyra+23}. We consider three different population of pebbles: the first is a power law distribution where the density of pebbles follow the power law in \eqs{eq:powerlaw}{eq:q1}, with the largest grain being pure ice and the smallest grain being pure silicate. The second is a bimodal distribution, given by \eqs{eq:secondmodel}{eq:q2}, where grains smaller than $1$\,mm are pure silicates, and larger grains follow a power law, with the largest grain being pure ice. The last model assumes all pebbles are silicates. 

We find the power law model does not reproduce the densities for high-mass KBOs. The model with purely silicate pebbles is able to reproduce the high density of Eris, yet fails to reproduce the densities of Haumea, Pluto, and Triton. The bimodal distribution, however, is capable of reproducing the densities of the dwarf planets but underestimates the density of Eris. Thus we speculate that Eris could have formed from a rock-rich filament or that it formed later in the solar system history when volatiles were lost through radial drift or photoevaporation. Nevertheless, it is conceivable that Eris formed with the density predicted our bimodal distribution, and subsequently lost ice through collisions, which we do not model. 

We find that the masses of KBOs are best reproduced between 15 and 22\,AU. Beyond this range, accretion rates are far too low to achieve dwarf-planet mass by the end of the disk lifetime. Conversely, inwards of 15\,AU, accretion rates are too high, resulting in masses that are orders of magnitudes larger than Triton and Pluto. 

Our solution avoids the timing problem that KBOs formed too early would melt and become compact, due to the energy released by $^{26}$Al. In our model, the first planetesimals are icy, and $^{26}$Al is mostly incorporated in the long phase of silicate pebble accretion, when most $^{26}$Al has already decayed. While the specific results on location of formation and final mass are dependent on the disk model adopted, the premise and conclusions of the work, namely the need to separate the silicates from ices and then preferentially accrete silicates, and that Bondi accretion and ice desorption from small grains are a way to accomplish these, are independent of the particular disk model.

Our results lend further credibility to the streaming instability as a planetesimal formation mechanism and to pebble accretion as a mechanism by which planetesimals grow into larger bodies. This model also provides yet another challenge for the previously held idea of planetesimal accretion as the main driver for growth. Growth through binary planetesimal accretion would result in planetesimals with similar densities regardless of size, with a maximum increase of a factor of two due to gravitational compaction. We show that multi-species streaming instability could result in planetesimals of nearly uniform composition and that a polydisperse pebble accretion model can have a significant impact on the final composition of a planetary body.

\acknowledgments

MHC, WL, DS, JBS, OMU, CCY, and ANY acknowledge support from the NASA Theoretical and Computational Astrophysical Networks (TCAN) via grant 80NSSC21K0497. MHC and WL are further supported by grant
\#80NSSC22K1419 from the NASA Emerging Worlds program, DS by NSF via grant AST-2007422, and CCY acknowledges the support from NASA via the Astrophysics Theory Program (grant \#80NSSC21K0141) and the Emerging Worlds program (grant \#80NSSC20K0347 and \#80NSSC23K0653). We acknowledge suggestions and constructive criticism from Anders Johansen, Zsolt S\'andor, Octavio Guilera, and Paul Estrada. The simulations presented in this paper utilized the Stampede cluster of the Texas Advanced Computing Center (TACC) at The University of Texas at Austin, through XSEDE/Access grant TG-AST140014, and the Discovery cluster at New Mexico State University \citep{TrecakovVonWolff21}. This work utilized resources from the New Mexico State University High Performance Computing Group, which is directly supported by the National Science Foundation (OAC-2019000), the Student Technology Advisory Committee, and New Mexico State University and benefits from inclusion in various grants (DoD ARO-W911NF1810454; NSF EPSCoR OIA-1757207; Partnership for the Advancement of Cancer Research, supported in part by NCI grants U54 CA132383 (NMSU)). This research was made possible by the open-source projects \texttt{jupyter}\ \citep{Jupyter}, \texttt{IPython}
\citep{IPython}, \texttt{matplotlib} \citep{matplotlib1, matplotlib2}, \texttt{NumPy} \citep{numpy}, \texttt{SymPy} \citep{sympy}, and \texttt{AstroPy} \citep{astropy}. 

\appendix

\section{Drag force backreaction with multiple species}
\label{app:dragforce}

\eqs{eq:position-update}{eq:eqdust} are solved for every particle $k$ in the simulation. The drag force accelerating particle $k$ is 

\beq
\v{f}_{d,k} = -\frac{\v{v}_k- \overline{\v{u}}(\v{x}_k)}{\tau_k}
\eeq

\noindent where $\tau$, the friction time, is the timescale at which the relative velocity between the gas and particle $k$ is reduced by a factor of $e$. For grains smaller than the mean free path of the gas \citep[Epstein drag, ][]{Epstein1924}, the appropriate regime for our investigation \citep[e.g.][]{Johansen+14}, this timescale is related to the grain size by

\begin{equation}\label{eq:tau}
    \tau_k = \frac{a_k \ \rho_{\bullet, k}}{v_{\rm th}\overline{\rho_g}(\v{x}_k)}.
\end{equation}

\noindent Here $a$ is the grain radius, $\rho_{\bullet}$ is the grain internal density and $v_{\rm th} = \sqrt{8/\pi}c_s$ is the thermal speed of the molecules. The friction time is often non-dimensionalized by defining the Stokes number $\St$,

\begin{equation}\label{eq:st}
    \St_k \equiv \varOmega_0 \tau_k.
\end{equation}

The quantities $\overline{\v{u}}(\v{x}_k)$ and $\overline{\rho_g}(\v{x}_k)$ refer to the gas velocity $\v{u}(\v{x})$ and density $\rho_g (\v{x})$, defined on the Eulerian mesh positions $\v{x}$, interpolated to the position $\v{x}_k$ of particle $k$. The interpolation for a quantity $\psi$ is done as described in \cite{YoudinJohansen07} 

\beq
\overline{\psi}(\v{x}_k) = \sum W \left(\v{x}_k - \v{x}\right) \psi (\v{x})
\label{eq:kernel-interp}
\eeq

\noindent where the sum is done over mesh positions. The mesh weight kernel $W$ is chosen as the Triangular Shaped Cloud algorithm \citep{HockneyEastwood88}, where the nearest 3 cells to the particle in each direction (27 cells in total) contribute to the interpolation. Incidentally, the pebble density needed for \eq{eq:poisson} is calculated on the mesh cells according to 

\beq
\rho_d(\v{x}) = \frac{1}{V(\v{x})}\sum_k W(\v{x}_k-\v{x}) m_k, 
\eeq

\noindent and once the selfgravity potential $\varPhi(\v{x})$ is calculated and the gradient $\grad{\varPhi}(\v{x})$ taken on the mesh, the interpolation $\overline{\grad{\varPhi}}(\v{x}_k)$ to the position of each particle $k$ is done according to \eq{eq:kernel-interp} and added to \eq{eq:eqdust}. 

The drag force backreaction $\v{f}_g$ onto a mesh cell centered at $\v{x}$ is calculated in a momentum-conserving way, via Newton's 3rd law \citep{YoudinJohansen07}

\beq
\v{f}_{g}(\v{x}) = -\frac{1}{\rho_g (\v{x}) V (\v{x})} \sum_k W \left(\v{x}_k  - \v{x}\right)m_k\,f_{d,k}
\eeq

\noindent where $V (\v{x})$ is the mesh cell volume and $m_k$ the mass of particle $k$. In this formulation, the dragforce backreaction for multiple particle species is straightforward, as the sum collects the individual contribution of each particle $k$. 

\section{Heating from decay of $^{\rm26}$Al}
\label{app:heating}

\begin{figure}
    \begin{center}
    \resizebox{.5\textwidth}{!}{\includegraphics{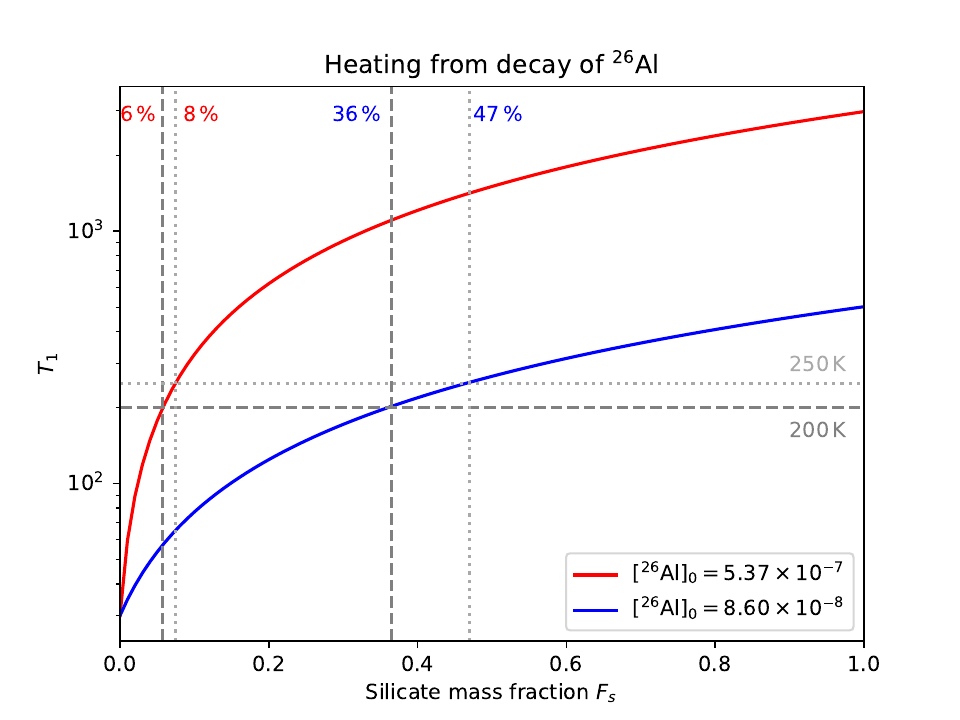}}
    \end{center}
    \caption{Body temperature $T_1$ resulting from heating from decay of $^{\rm26}$Al, as a function of mass fraction of silicates, $F_s$.}
\label{fig:csivst}
\end{figure}

We estimate here whether the small bodies produced in the streaming instability model would melt as a result of heating from $^{\rm26}$Al. Considering the model from \citet{Castillo-Rogez+09}, the volumetric heating rate $\mathcal{H}$ is

\begin{equation}
\mathcal{H} = \rho \, F_s \, {\rm [^{26}Al]}_0 \, \mathcal{H}_0 \, e^{-\lambda t}
\end{equation}

\noindent Where $\rho$ is the protoplanet's density, $F_s$ is the silicate mass fraction, ${\rm [^{26}Al]}_0$ is the initial isotopic abundance
of $^{\rm 26}$Al in ordinary chondrites, $\mathcal{H}_0$ is the heat production rate per mass, and $\lambda=\ln(2)/t_{1/2}$ is
the decay constant of $^{\rm 26}$Al; $t_{1/2}$ is the half-life. We find the energy by integrating in time and multiplying by the volume $V$

\begin{eqnarray}
  Q(t) &=& V \int_0^{t} \mathcal{H} (t^\prime) dt^\prime\\
  &=& M_p \, F_s \, {\rm [^{26}Al]}_0 \, \mathcal{H}_0 \, \lambda^{-1} \, \left(1 - e^{-\lambda t} \right)
\end{eqnarray}

\noindent where we substituted the protoplanet's mass $M_p=\rho V$. If we assume that all this energy is trapped within the body, we can find the temperature $T_1$ achieved by setting $t\rightarrow\infty$, and equating with the heat capacity equation

\beq
Q = M_p \,c_p \,\Delta T
\eeq

\noindent where $c_p$ is the heat capacity at constant pressure. Then solving for $\Delta T$

\beq
\Delta T = F_s \, {\rm [^{26}Al]}_0 \, \mathcal{H}_0 \, \lambda^{-1} \, c_p^{-1}
\eeq

\noindent Substituting the values $c_p = 2108 \ {\rm J\,kg^{-1}\,K^{-1}}$ \citep{Marquet15}, ${\rm [^{26}Al]}_0 = \left(8.6\times 10^{-8},5.37\times 10^{-7}\right)$ \citep{ParhiPrialnik23}, 
$\mathcal{H}_0 = 0.355 \ {\rm W/kg}$ \citep{Castillo-Rogez+09}, $t_{1/2} = 0.717 \ {\rm Myr}$, and $T_0 = 30\,{\rm K}$, we find the $F_s$ vs $T_1$ (the final temperature) relationship shown in \fig{fig:csivst}. The blue curve is for the lower value of ${\rm [^{26}Al]}_0$, the red curve for the higher value. The light grey
dotted line is for heating up to 250\,K; the dark grey dashed line for heating up to 200\,K. It seems
that according to this model, with no energy release, a temperature of 200\,K (250\,K) is achieved
at a mere $F_s$=0.06 (0.08) for the higher $^{\rm 26}$Al abundance. The lower abundance of $^{\rm 26}$Al reaches
200\,K for $F_s$=0.36, or 250\,K at $F_s$=0.47.

In this model, the higher abundance would not allow for porosity retention, but the lower
  abundance accomodates the $\gtrsim 65\%$ ice mass fraction seen in \fig{fig:mass-vs-ice} for the products of streaming instability. We
highlight that these are likely overestimates for the heating rate because, as noted by \citet{ParhiPrialnik23}, part of the heat is used to sublimate the hypervolatiles CO and CH$_4$, which
considerably lowers the attained final temperature. We note that this is consistent with the very
low densities of the small KBOs, without necessitating unrealistic porosities.

\bibliographystyle{apj}

\end{document}